\documentclass[a4paper,11pt]{article}
\pdfoutput=1
\usepackage[T1]{fontenc}
\usepackage{amsmath,amssymb}
\usepackage{epsfig}
\usepackage{graphicx}
\usepackage{jheppub} 

\title{\boldmath Charged Lepton Flavour Violation and Neutrinoless Double Beta Decay in Left-Right Symmetric Models with Type I+II Seesaw}
\author[a]{Debasish Borah}
\emailAdd{dborah@iitg.ernet.in}
\affiliation[a]{Department of Physics, Indian Institute of Technology Guwahati, Assam-781039, India}
\author[b]{Arnab Dasgupta}
\emailAdd{arnab.d@iopb.res.in}
\affiliation[b]{Institute of Physics, Sachivalaya Marg, Bhubaneshwar-751005, India}
\abstract{
We study the new physics contributions to neutrinoless double beta decay ($0\nu\beta \beta$) half-life and lepton flavour violation (LFV) amplitude within the framework of the minimal left-right symmetric model (MLRSM). Considering all possible new physics contributions to $0\nu\beta \beta$ and charged lepton flavour violation $\mu \rightarrow e \gamma, \mu \rightarrow 3e$ in MLRSM, we constrain the parameter space of the model from the requirement of satisfying existing experimental bounds. Assuming the breaking scale of the left-right symmetry to be $\mathcal{O}(1)$ TeV accessible at ongoing and near future collider experiments, we consider the most general type I+II seesaw mechanism for the origin of tiny neutrino masses. Choosing the relative contribution of the type II seesaw term allows us to calculate the right handed neutrino mass matrix as well as Dirac neutrino mass matrix as a function of the model parameters, required for the calculation of $0\nu\beta \beta$ and LFV amplitudes. We show that such a general type I+II seesaw structure results in more allowed parameter space compared to individual type I or type II seesaw cases considered in earlier works. In particular, we show that the doubly charged scalar masses $M_{\Delta}$ are allowed to be smaller than the heaviest right handed neutrino mass $M_N$ from the present experimental bounds in these scenarios which is in contrast to earlier results with individual type I or type II seesaw showing $M_{\Delta} > M_N$.
}
\begin{document}
\maketitle


\section{Introduction}
Observations of non-zero neutrino masses and mixing \cite{PDG, kamland08} has been one of the most compelling evidences of the existence of beyond standard model (BSM) physics. Although the recently observed Higgs boson is believed to be responsible for the masses of all the known fundamental particles, it can not account for observed neutrino masses due to the absence of any renormalizable couplings between the Higgs and neutrino fields. The recent neutrino experiments MINOS \cite{minos}, T2K \cite{T2K}, Double ChooZ \cite{chooz}, Daya-Bay \cite{daya} and RENO \cite{reno} have not only confirmed the earlier observations of tiny neutrino masses, but also measured the neutrino parameters more precisely. The $3\sigma$ global fit values of neutrino oscillation parameters that have appeared in the recent analysis of \cite{schwetz14}  and \cite{valle14}  are shown in table \ref{tab:data1}.
\begin{center}
\begin{table}[htb]
\begin{tabular}{|c|c|c|c|c|}
\hline
Parameters & NH \cite{schwetz14} & IH \cite{schwetz14} & NH \cite{valle14} & IH \cite{valle14} \\
\hline
$ \frac{\Delta m_{21}^2}{10^{-5} \text{eV}^2}$ & $7.02-8.09$ & $7.02-8.09 $ & $7.11-8.18$ & $7.11-8.18 $ \\
$ \frac{|\Delta m_{31}^2|}{10^{-3} \text{eV}^2}$ & $2.317-2.607$ & $2.307-2.590 $ & $2.30-2.65$ & $2.20-2.54 $ \\
$ \sin^2\theta_{12} $ &  $0.270-0.344 $ & $0.270-0.344 $ &  $0.278-0.375 $ & $0.278-0.375 $ \\
$ \sin^2\theta_{23} $ & $0.382-0.643$ &  $0.389-0.644 $ & $0.393-0.643$ &  $0.403-0.640 $ \\
$\sin^2\theta_{13} $ & $0.0186-0.0250$ & $0.0188-0.0251 $ & $0.0190-0.0262$ & $0.0193-0.0265 $ \\
$ \delta $ & $0-2\pi$ & $0-2\pi$ & $0-2\pi$ & $0-2\pi$ \\
\hline
\end{tabular}
\caption{Global fit $3\sigma$ values of neutrino oscillation parameters \cite{schwetz14, valle14}.}
\label{tab:data1}
\end{table}
\end{center}
Although the $3\sigma$ range for the leptonic Dirac CP phase $\delta$ is $0-2\pi$, there are two possible best fit values of it found in the literature: $306^o$ (NH), $254^o$ (IH) \cite{schwetz14} and $254^o$ (NH), $266^o$ (IH) \cite{valle14}. There has also been a hint of this Dirac phase to be $-\pi/2$ as reported by \cite{diracphase} recently. Although the absolute mass scale of the neutrinos are not yet known, we have an upper bound on the sum of absolute neutrino masses from cosmology, given by the Planck experiment $\sum_i \lvert m_i \rvert < 0.23$ eV \cite{Planck13}. This bound has become even more strict $\sum_i \lvert m_i \rvert < 0.17$ eV from the latest analysis by Planck collaboration \cite{Planck15}.

The easiest way to account for non-zero neutrino masses is to introduce at least two right handed neutrinos into the standard model (SM). This will allow a Dirac coupling between neutrino and the Higgs, similar to other fermions in the SM. However, the corresponding Yukawa couplings have to be very small (around $10^{-12}$) in order to generate neutrino mass of order $0.1$ eV. Such highly unnatural fine-tuned values suggest a richer dynamical mechanism behind the origin of tiny but non-zero neutrino masses. This type of fine-tuning can be avoided in the so called seesaw mechanisms of neutrino masses, the most popular BSM framework explaining the origin of neutrino mass. Although seesaw mechanisms can be implemented in a variety of ways, the basic idea is to introduce additional fermionic or scalar fields heavier than the electroweak scale, such that the tiny neutrino masses result from the hierarchy between electroweak and seesaw scale. Such seesaw mechanisms broadly fall into three categories namely, type I \cite{ti}, type II \cite{tii0,tii} and type III \cite{tiii}. These generic seesaw mechanisms give rise to tiny neutrino masses of Majorana type by introducing new interactions with lepton number violation (LNV) through heavy fields. The same heavy fields can also give rise to lepton flavour violation (LFV) in the charged fermion sector. Therefore, these seesaw mechanisms offer different possible ways for experimental verification, from discovery machines like the Large Hadron Collider (LHC) to low energy experiments looking for LFV, LNV signals. Some earlier references on such LHC searches can be found in \cite{lrlhc1, ndbd00}. Such models are expected to undergo further scrutiny at other particle collider experiments which are being planned at present. Some recent works discussing the sensitivity and discovery potential of experiments like the Future Circular Collider (FCC), the Circular Electron Positron Collider - Super Proton-Proton Collider (CEPC/SppC), the International Linear Collider (ILC) and the Compact Linear Collider (CLIC) to similar new physics effects can be found at \cite{futureCollider, futureCollider2}. In the present work, we consider the latter possibility as a probe of these seesaw models. In particular, we study the possibility of observable signatures at experiments looking for charged lepton flavour violation like $\mu^- \rightarrow e^- e^- e^+, \mu^- \rightarrow e^- \gamma$ and lepton number violating processes like neutrinoless double beta decay, often referred to as $0\nu\beta \beta$ where a heavier nucleus decays into a lighter one and two electrons $(A, Z) \rightarrow (A, Z+2) + 2e^- $. For a review on $0\nu\beta \beta$, please refer to \cite{NDBDrev}. The strength of LFV processes in the SM remain suppressed much below the sensitivity of experiments \cite{sindrum, MEG, MEG2} due to the smallness of neutrino mass. Similarly, the SM contribution to $0\nu \beta \beta$ also remains much below the current experimental bounds \cite{kamland, GERDA, kamland2} unless the lightest neutrino mass falls in the quasi-degenerate regime, which is already disfavored by Planck data \cite{Planck13, Planck15}. However, in the presence of additional new particles around the TeV corner, current as well as future experiments can be sensitive to such processes. Here we consider TeV scale type I and type II seesaw as the origin of neutrino mass and study the consequences for LFV and LNV processes. We study them within the framework of minimal left-right symmetric model (MLRSM) \cite{lrsm, lrsmpot} which implements these two seesaw mechanisms naturally. This model which can be realised within the framework of grand unified theories like $SO(10)$ also relates the origin of neutrino mass to the spontaneous breaking of parity. Several earlier works \cite{tii0, NDBDprev} have calculated the new physics contributions to $0\nu \beta \beta$ within the framework of MLRSM. More recently, the authors of \cite{ndbd00, ndbd0} studied the new physics contributions to $0\nu \beta \beta$ process for TeV scale MLRSM with dominant type II seesaw. There have also been several works \cite{ndbd1, ndbd10, ndbd101, ndbd102} where type I seesaw limit was also included into the computation of $0\nu \beta \beta$ in MLRSM. Some more detailed analyses incorporating left-right gauge mixing were discussed in the works \cite{ndbd2, ndbd21}. Recently, some more works appeared connecting lepton number violation responsible for $0\nu \beta \beta$ with collider observables \cite{lnvcollider}. In particular, MLRSM and heavy neutrinos have been studied with respect to the Large Hadron Electron Collider (LHeC) in \cite{LHeC}.

In almost all the works discussing LFV and $0\nu \beta \beta$ in MLRSM, calculations were done by assuming either type I or type II seesaw dominance at a time. It is therefore straightforward to relate the parameters involved in either type I or type II seesaw term directly with the light neutrino ones. However, if both the seesaw terms are sizeable then one has more freedom to tune the individual seesaw terms in a way that their combination gives the effective light neutrino masses and mixing. In a recent work \cite{ndbd101}, we considered equally dominant type I and type II seesaw, with the type I seesaw mass matrix possessing a $\mu-\tau$ symmetry, or, more specifically, Tri-Bimaximal or TBM type mixing. We then studied the new physics contributions to $0\nu \beta \beta$ amplitude by taking experimental constraints on LFV process $\mu \rightarrow 3e$, masses of triplet scalars, new gauge bosons and right handed neutrinos. In another recent work \cite{ndbd102}, scalar triplet contributions to LFV processes $\mu \rightarrow 3e, \mu \rightarrow e \gamma$ as well as $0\nu \beta \beta$ were studied for either type I or type II dominant cases. The authors showed that the current experimental bounds still allow light scalar triplet mass in MLRSM which was earlier thought to be around ten times heavier than the heaviest right handed neutrino mass \cite{ndbd00}. To be more specific, the authors of \cite{ndbd102} showed that for heaviest right handed neutrino mass as low as 400 GeV, the triplet scalars are allowed to be as low as around 800 GeV for right handed charged gauge boson mass 3.5 TeV. Here we extend both these works \cite{ndbd101, ndbd102} by considering more general type I and type II seesaw terms with comparable strength and study their implications in LFV processes $\mu \rightarrow 3e, \mu \rightarrow e \gamma$ and LNV process like $0\nu \beta \beta$. Instead of considering any specific mass matrix structure for either type I or type II seesaw mass matrix, we consider a very general mass matrix for one of the seesaw terms. The other seesaw mass matrix then gets automatically fixed from the neutrino mass formula by demanding agreement with light neutrino data. We call it democratic type I - type II seesaw scenario. One can also assume some specific structure of one of these mass matrices as was done in \cite{ndbd101} to reduce the number of free parameters. However, in the absence of additional flavour symmetries, such realisations are ad-hoc to some extent and hence we intend to do a more general study in this work.

This paper is organised as follows. In section \ref{sec1}, we first briefly discuss the left-right symmetric model and then summarise the origin of neutrino masses in this model in subsection \ref{sec2}. In subsection \ref{sec1a}, we briefly point out the possible new physics sources to neutrinoless double beta decay amplitude. In section \ref{sec:lfvlhc} we briefly discuss charged lepton flavor violation in the model and then comment on the existing collider constraints in subsection \ref{sec:collider}. In section \ref{sec:type12}, we outline the details of type I+II seesaw structure. In section \ref{sec3}, we discuss our numerical analysis and finally conclude in \ref{sec4}.

\section{Minimal Left-Right Symmetric Model}
\label{sec1}
Left-Right Symmetric Model \cite{lrsm, lrsmpot} is one of the best motivated BSM frameworks which is based on the idea that Nature is parity symmetric at high energy scale and low energy parity violation in electroweak interactions occurs due to spontaneous breaking of parity. The model is made parity symmetric by extending the gauge symmetry of the SM from $SU(3)_c \times SU(2)_L \times U(1)_Y$ to $SU(3)_c \times SU(2)_L \times SU(2)_R \times U(1)_{B-L}$ such that the right handed fermions have similar $SU(2)$ gauge interactions with equal strength $g_R = g_L$. The $U(1)_{B-L}$ gauge anomaly cancellation conditions require the inclusion of right handed neutrinos as part of $SU(2)_R$ fermion doublets. This ensures the presence of seesaw mechanism as origin of light neutrino masses. The right handed neutrinos responsible for type I seesaw as well as the additional gauge bosons acquire heavy masses when the enhanced gauge symmetry of the model $SU(2)_R \times U(1)_{B-L}$ is broken down to the $U(1)_Y$ of SM by the vacuum expectation value (vev) of additional Higgs scalar, transforming as triplet under $SU(2)_R$ and having non-zero $U(1)_{B-L}$ charge. The left handed Higgs triplet on the other hand, can give tiny Majorana masses to the SM neutrinos through type II seesaw mechanism.

The fermion content of the MLRSM is
\begin{equation}
Q_L=
\left(\begin{array}{c}
\ u_L \\
\ d_L
\end{array}\right)
\sim (3,2,1,\frac{1}{3}),\hspace*{0.8cm}
Q_R=
\left(\begin{array}{c}
\ u_R \\
\ d_R
\end{array}\right)
\sim (3^*,1,2,\frac{1}{3}),\nonumber 
\end{equation}
\begin{equation}
\ell_L =
\left(\begin{array}{c}
\ \nu_L \\
\ e_L
\end{array}\right)
\sim (1,2,1,-1), \quad
\ell_R=
\left(\begin{array}{c}
\ \nu_R \\
\ e_R
\end{array}\right)
\sim (1,1,2,-1) \nonumber
\end{equation}
Similarly, the Higgs content of the minimal LRSM is
\begin{equation}
\Phi=
\left(\begin{array}{cc}
\ \phi^0_{11} & \phi^+_{11} \\
\ \phi^-_{12} & \phi^0_{12}
\end{array}\right)
\sim (1,2,2,0)
\nonumber 
\end{equation}
\begin{equation}
\Delta_L =
\left(\begin{array}{cc}
\ \delta^+_L/\surd 2 & \delta^{++}_L \\
\ \delta^0_L & -\delta^+_L/\surd 2
\end{array}\right)
\sim (1,3,1,2), \hspace*{0.2cm}
\Delta_R =
\left(\begin{array}{cc}
\ \delta^+_R/\surd 2 & \delta^{++}_R \\
\ \delta^0_R & -\delta^+_R/\surd 2
\end{array}\right)
\sim (1,1,3,2) \nonumber
\end{equation}
Here the numbers in brackets denote the transformations of respective fields under the gauge symmetry of the model that is, $SU(3)_c\times SU(2)_L\times SU(2)_R \times U(1)_{B-L}$. This gauge symmetry gets broken down to the symmetry of the standard model when the neutral component of the Higgs triplet $\Delta_R$ acquires a vev at a high energy scale. Consequently, the symmetry of the SM gets broken down to the $U(1)$ of electromagnetism by the vev of the neutral component of Higgs bidoublet $\Phi$:
$$ SU(2)_L \times SU(2)_R \times U(1)_{B-L} \quad \underrightarrow{\langle
\Delta_R \rangle} \quad SU(2)_L\times U(1)_Y \quad \underrightarrow{\langle \Phi \rangle} \quad U(1)_{em}$$
The symmetry breaking of $SU(2)_R \times U(1)_{B-L}$ into the $U(1)_Y$ of standard model can also be achieved at two stages by choosing a non-minimal scalar sector which for example, was shown in\cite{lrdb}.

\subsection{Neutrino Mass in MLRSM}
\label{sec2}
The gauge symmetry of the MLRSM allows the following Yukawa terms relevant for tiny neutrino masses can be written in Weyl spinor notations as,
\begin{eqnarray}
{\cal L}^{II}_\nu &=& y_{ij} \ell_{iL} \Phi \ell_{jR}+ y^\prime_{ij} \ell_{iL}
\tilde{\Phi} \ell_{jR} +h.c.
\nonumber \\
&+& f_{ij}\ \left(\ell_{iR}^T \ C \ i \sigma_2 \Delta_R \ell_{jR}+
(R \leftrightarrow L)\right)+h.c.
\label{treeY}
\end{eqnarray}
where $\tilde{\Phi} = \tau_2 \Phi^* \tau_2$. In the above Yukawa Lagrangian, the indices $i, j = 1, 2, 3$ correspond to the three generations of fermions. The Majorana Yukawa couplings $f$ are the same for both left and right handed neutrinos
because of the in built left-right symmetry $(f_L = f_R)$. These couplings $f$ give rise to the Majorana mass terms of both left handed and right handed neutrinos after the triplet Higgs fields $\Delta_{L,R}$ acquire non-zero vev. These mass terms appear in the seesaw formula of MLRSM that can be written as
\begin{equation}
M_{\nu}=M_{\nu}^{II} + M_{\nu}^I
\label{type2a}
\end{equation}
where the usual type I seesaw term $M_{\nu}^I$ is given by the expression,
\begin{equation}
M_{\nu}^I=-m_{LR}M_{RR}^{-1}m_{LR}^{T}.
\end{equation}
Here  $m_{LR} = (y v_1 + y^\prime v_2)/\sqrt{2}$ is the Dirac neutrino mass matrix, with $v_{1,2}$ are the vev's of the neutral components of the Higgs bidoublet. It is worth mentioning that in the framework of MLRSM, $M_{RR}$ arises naturally as a result of left-right symmetry breaking at high energy scale and it appears both in type I and type II seesaw terms. In MLRSM, $M_{RR}$ can be expressed as $M_{RR}=\sqrt{2}v_{R}f_{R}$. The first term $M_{\nu}^{II}$ in equation (\ref{type2a}) is due to the vev of $SU(2)_{L}$ Higgs triplet. Thus, it can be written as $M_{\nu}^{II}=\sqrt{2}f_{L}v_{L}$ in a way similar to $M_{RR}=\sqrt{2}f_{R}v_{R}$, where $v_{L,R}$ denote the vev's and $f_{L,R}$ are symmetric $3\times3$ matrices. The left-right symmetry demands $f_{R}=f_{L}=f$ as mentioned above. The induced vev for the left-handed triplet $v_{L}$ can be shown for MLRSM to be
$$v_{L}=\gamma \frac{M^{2}_{W_L}}{v_{R}}$$
with $M_{W_L}\sim 80.4$ GeV being the charged electroweak vector boson mass and $v_R$ being the high energy scale at which left-right symmetry gets broken spontaneously such that 
$$ |v_{L}|<<M_{W_L}<<|v_{R}| $$ 
In general, $\gamma$ is a dimensionless parameter which can be written in terms of the vev's $v_1, v_2$ and several dimensionless couplings in the scalar potential of MLRSM.
Without any fine tuning $\gamma$ is expected to be of the order unity ($\gamma\sim 1$) following the results from Deshpande et al. \cite{lrsmpot}. However, for TeV scale type I+II seesaw, $\gamma$ has to be fine-tuned as we discuss later. The type II seesaw formula in equation (\ref{type2a}) can now be expressed as
\begin{equation}
M_{\nu}=\gamma (M_{W_L}/v_{R})^{2}M_{RR}-m_{LR}M^{-1}_{RR}m^{T}_{LR}
\label{type2}
\end{equation}
\begin{figure}[!h]
\centering
\begin{tabular}{ccc}
\epsfig{file=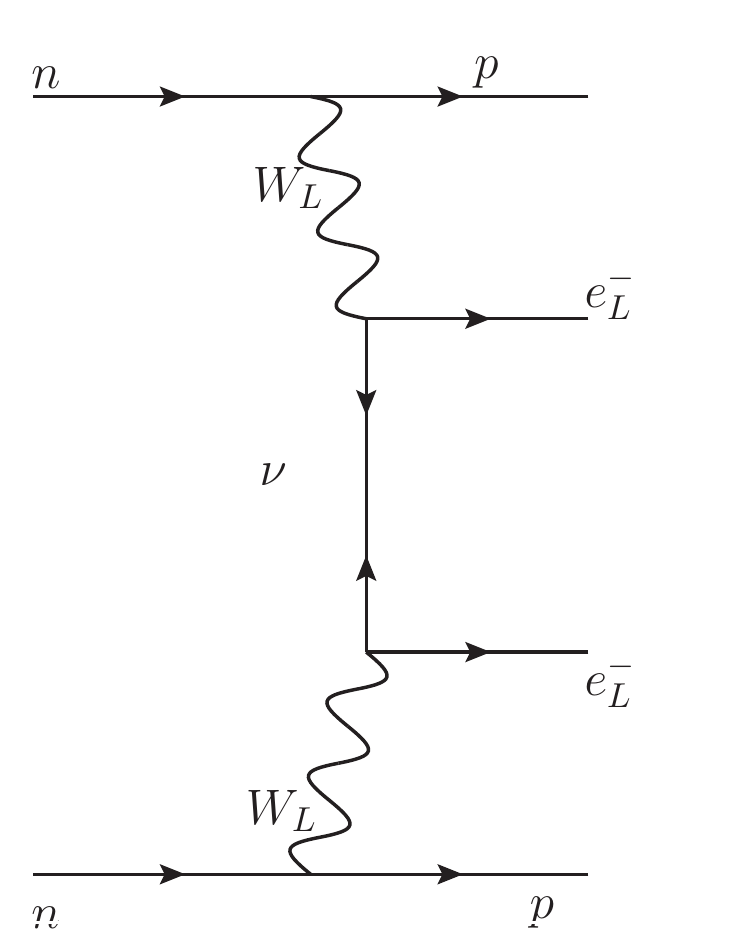,width=0.3\textwidth,clip=}&
\epsfig{file=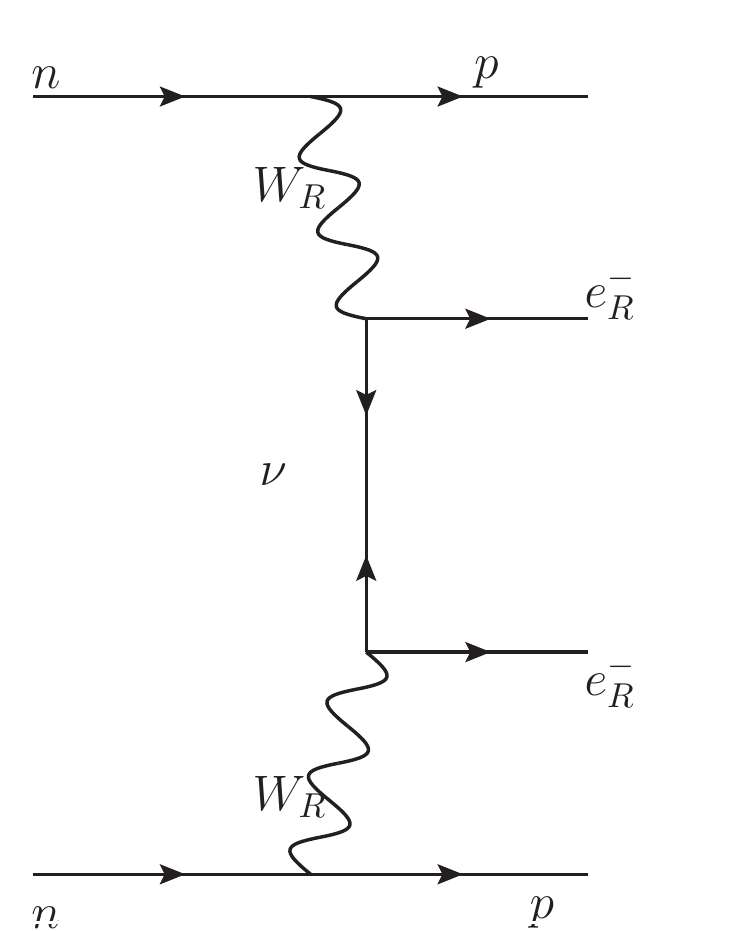,width=0.3\textwidth,clip=} &
\epsfig{file=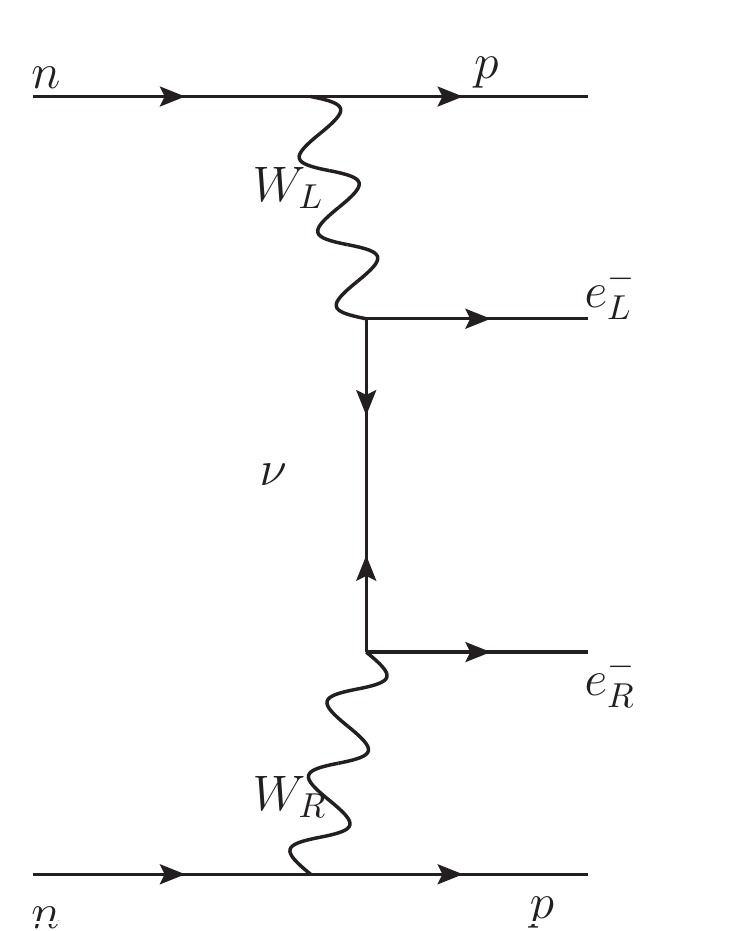,width=0.3\textwidth,clip=}
\end{tabular}
\caption{Feynman diagrams for Neutrinoless double beta decay due to $\nu_L-W_L-W_L, \nu_L-W_R-W_R, \nu_L-W_L-W_R$ contributions.}
\label{fig0}
\end{figure}
\begin{figure}[!h]
\centering
\begin{tabular}{ccc}
\epsfig{file=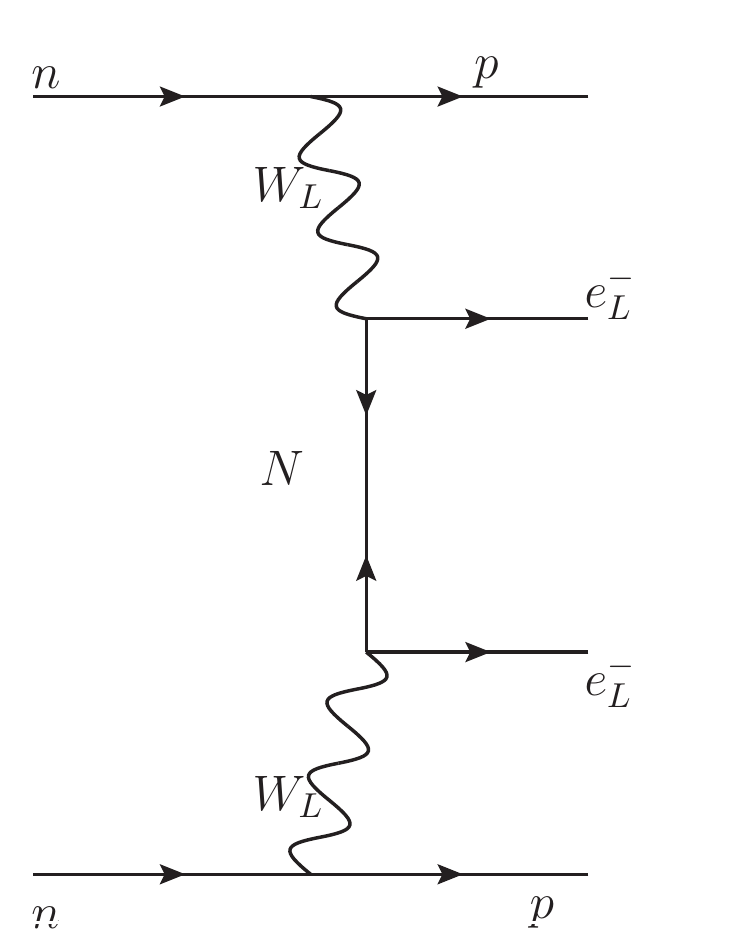,width=0.3\textwidth,clip=} &
\epsfig{file=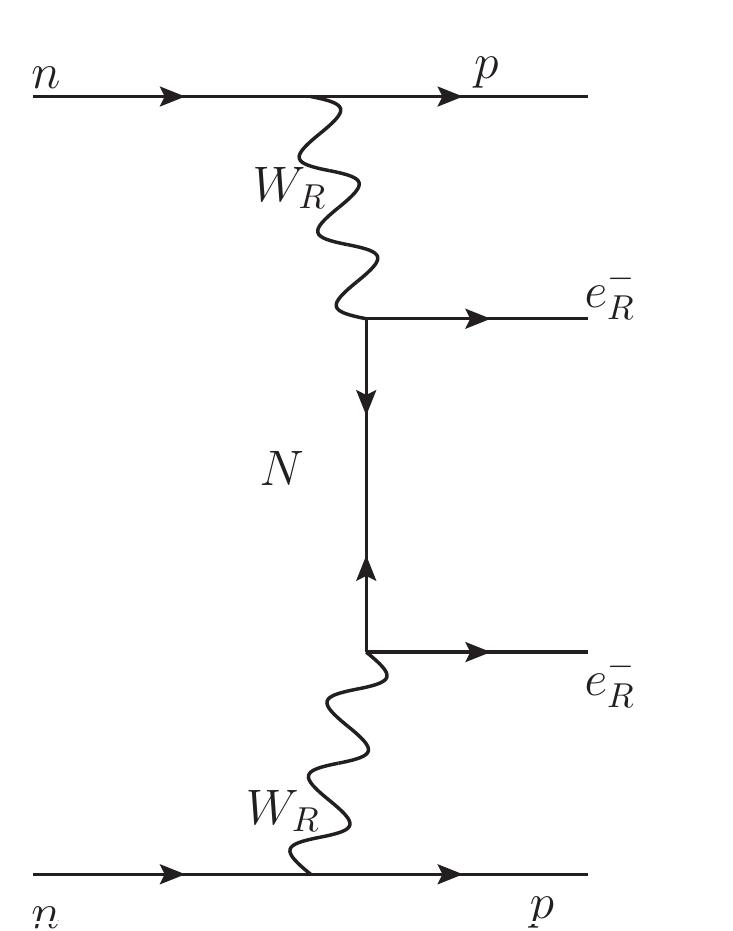,width=0.3\textwidth,clip=} &
\epsfig{file=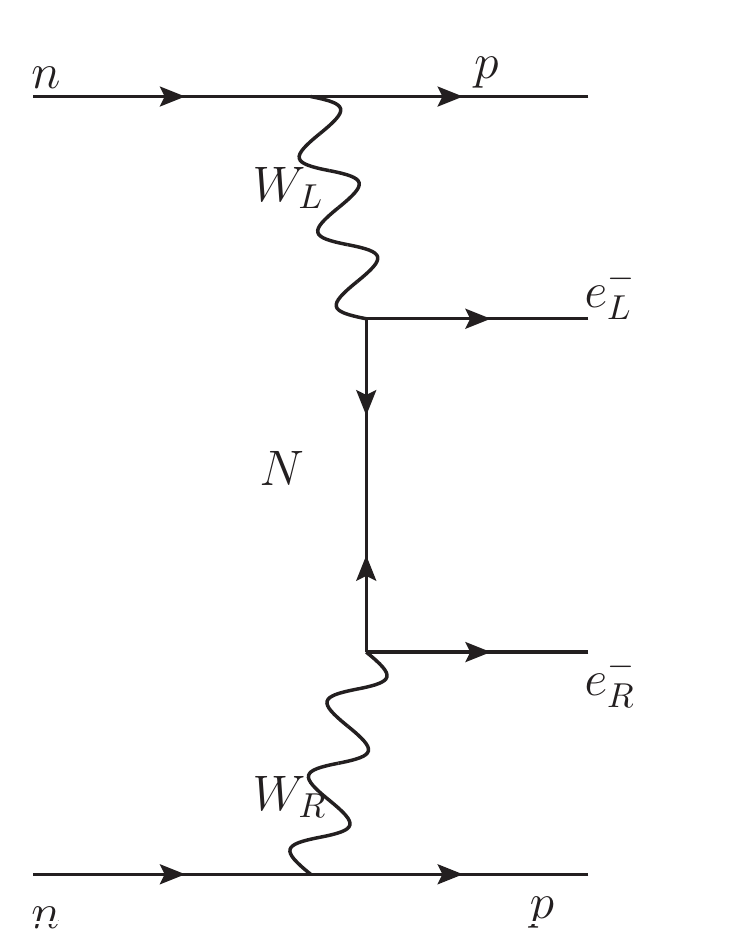,width=0.3\textwidth,clip=} 
\end{tabular}
\caption{Feynman diagrams for Neutrinoless double beta decay due to $\nu_R-W_L-W_L, \nu_R-W_R-W_R, \nu_R-W_L-W_R$ contributions.}
\label{fig01}
\end{figure}

\begin{figure}[!h]
\centering
\begin{tabular}{ccc}
\epsfig{file=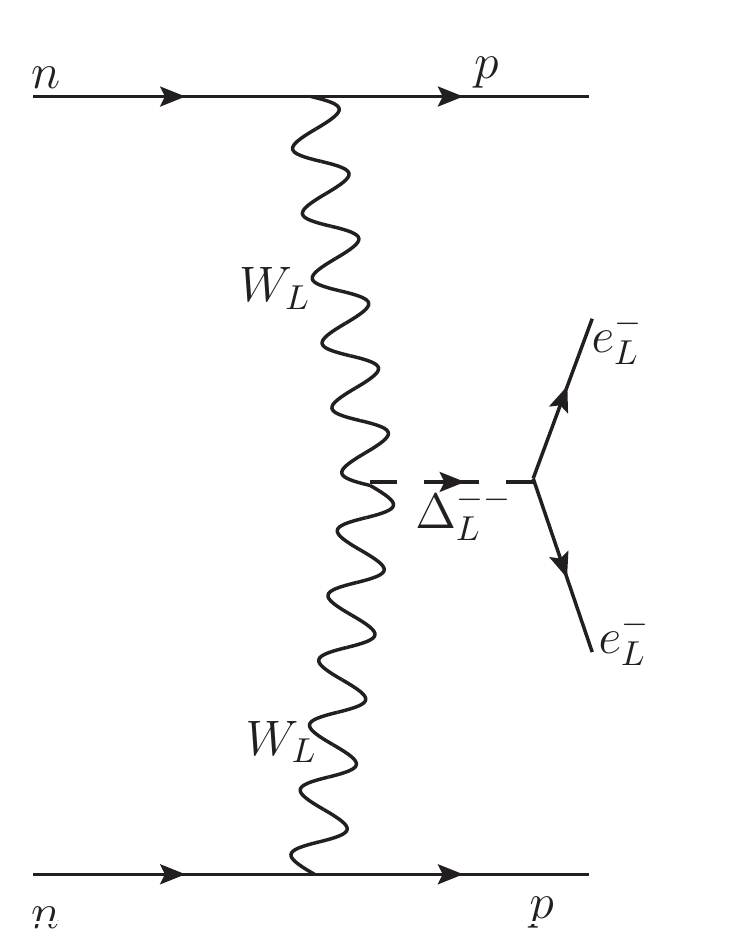,width=0.3\textwidth,clip=}&
\epsfig{file=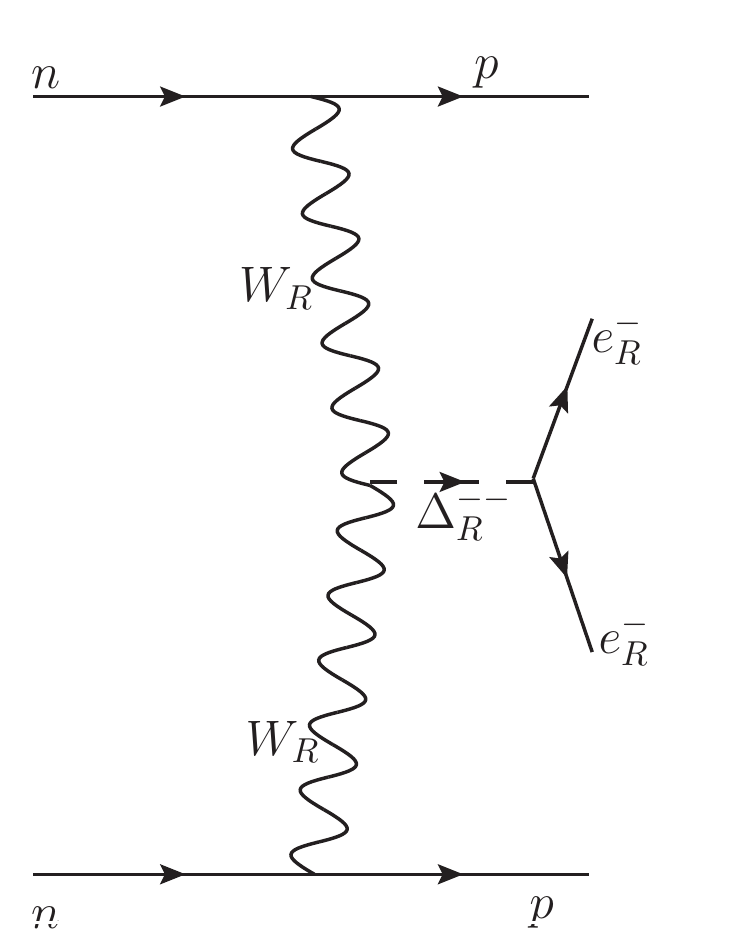,width=0.3\textwidth,clip=} &
\epsfig{file=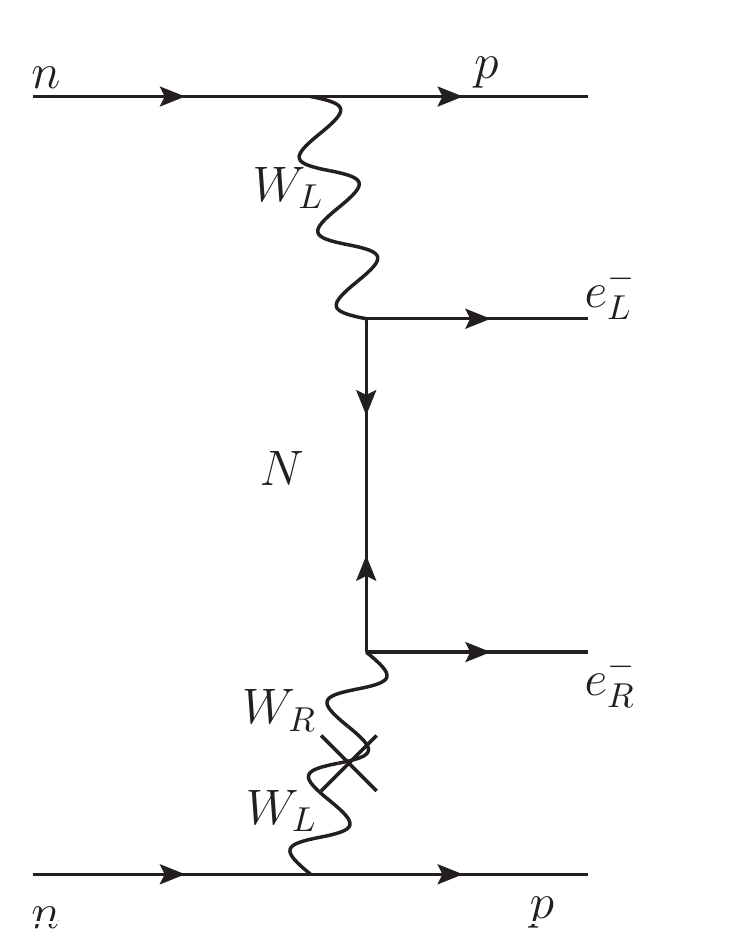,width=0.3\textwidth,clip=}
\end{tabular}
\caption{Feynman diagrams for Neutrinoless double beta decay due to $\Delta_{L,R}$ and $W_L-W_R$ mixing contributions.}
\label{fig02}
\end{figure}

\subsection{$0\nu \beta \beta$ in MLRSM}
\label{sec1a}
As the MLRSM contains several new fields which are not present in the SM, there can enhancement to neutrinoless double beta decay and charged lepton flavour violation amplitude. The corresponding Feynman diagrams given in earlier works, for example \cite{ndbd1} have been reproduced here, as shown in figure \ref{fig0}, \ref{fig01}, \ref{fig02} including the one with the standard light neutrino contribution.  The complete list of MLRSM contributions to $0\nu \beta \beta$ can be listed as follows:
\begin{enumerate}
\item The light neutrino contribution comes from the Feynman diagram where the intermediate particles are $W_L$ bosons and light neutrinos. The amplitude of this process depends upon the leptonic mixing matrix elements and the light neutrino masses. This corresponds to the first diagram in figure \ref{fig0}.
\item The light neutrino contribution can come from the Feynman diagram mediated by $W_R$ bosons such that the interaction between light neutrinos and $W_R$ boson is proportional to the mixing between light and heavy neutrinos. This corresponds to the second diagram in figure \ref{fig0}. Such a mixing between light and heavy neutrinos is usually suppressed from the constraints on non-unitarity of the leptonic mixing matrix \cite{nonUnitary}.
\item The light neutrino contribution can also come from the Feynman diagram mediated by both $W_L$ and $W_R$. The amplitude depends upon the mixing between light and heavy neutrinos, leptonic mixing matrix elements, light neutrino masses and $W_R$ mass. This is shown as the third diagram in figure \ref{fig0}.
\item The heavy right handed neutrino $\nu_R$ contribution can come from the Feynman diagrams mediated by $W_L$ bosons such that the interaction between heavy neutrinos and $W_L$ boson is suppressed by the mixing between light and heavy neutrinos. This is shown in the first panel of figure \ref{fig01}. 
\item The dominant heavy right handed neutrino contribution comes from the Feynman diagrams mediated by $W_R$ boson. The corresponding amplitude depends upon the elements of right handed leptonic mixing matrix and masses of $\nu_R$. This corresponds to the second diagram in figure \ref{fig01}.
\item The heavy right handed neutrino contribution can come from the Feynman diagram where the intermediate particles are $W_L$ and $W_R$ simultaneously. The amplitude depends upon the right handed leptonic mixing elements, mixing between light and heavy neutrinos as well as heavy neutrino masses. This is the third diagram in figure \ref{fig01}.
\item The triplet Higgs scalars $\Delta_L$ and $\Delta_R$ can also contribute to neutrinos double beta decay through $W_L$ and $W_R$ mediation respectively. The amplitude depends upon the masses of $\Delta_{L,R}$ scalars as well as their couplings to leptons. These corresponds to first and second diagrams in figure \ref{fig02}.
\item Heavy neutrino contribution can also come from the Feynman diagram with $W_L-W_R$ mixing as shown in the third panel of figure \ref{fig02}. Such $W_L-W_R$ mixing is usually suppressed by electroweak precision data as well as direct searches at colliders. Using the limits from direct searches for the same-sign dilepton signal at the LHC \cite{rhnlhc}, the authors of reference \cite{ndbd21} estimated such a mixing to be $\leq 7.7 \times 10^{-4}$.
\end{enumerate}

The amplitude of the light neutrino contribution (first Feynman diagram in figure \ref{fig0}) considered here is 
\begin{equation}
A_{\nu L L} \propto G^2_F \sum_i \frac{m_i U^2_{ei}}{p^2} 
\end{equation}
with $p$ being the average momentum exchange for the process. In the above expression, $m_i$ are the masses of light neutrinos for $i=1,2,3$. $G_F = 1.17 \times 10^{-5} \; \text{GeV}^{-2}$ is the Fermi coupling constant and $U$ is the light neutrino mixing matrix. In fact, this mixing matrix $U$ is a part of the full $6\times 6$ mixing matrix, including heavy and light neutrinos. This mixing matrix can be written in terms of $3\times3$ matrices $U, V, S, T$ as
\begin{equation}
\left(\begin{array}{cc}
\ U & S \\
\ T & V
\end{array}\right) = \left(\begin{array}{cc}
\ 1-\frac{1}{2}R R^{\dagger} & R \\
\ -R^{\dagger} & 1-\frac{1}{2}R^{\dagger}R 
\end{array}\right) \left(\begin{array}{cc}
\ U_L & 0 \\
\ 0 & U_R
\end{array}\right)
\label{mixingmatrix}
\end{equation}
such that $U_L, U_R$ are the diagonalising matrices of light and heavy neutrino mass matices $M_{\nu}, M_{RR}$ respectively. Here $R=m_{LR} M^{-1}_{RR}$. Simplifying the above equation gives rise to 
$$ U=U_L - \frac{1}{2} R R^{\dagger} U_L, \;\;\;\; S= R U_R$$
$$T = -R^{\dagger} U_L, \;\;\;\; V = U_R - \frac{1}{2} R^{\dagger}R U_R$$

The contribution from $W^-_R, \Delta_R$ exchange (third Feynman diagram in figure \ref{fig0}) is given by the amplitude
\begin{equation}
A_{R\Delta} \propto G^2_F \left ( \frac{M_{W_L}}{M_{W_R}} \right )^4 \sum_i \frac{V^2_{ei} M_i}{M^2_{\Delta^{++}_R}}
\end{equation}
where $M_i$ are the masses of right handed neutrinos for $i=1,2,3$. There exists a mirror diagram similar to this where $W^-_R, \Delta_R$ are replaced by $W^-_L, \Delta_L$ and the corresponding amplitude is given by
\begin{equation}
A_{L\Delta} \propto G^2_F   \frac{(M_{\nu}^{II})_{ee}}{M^2_{\Delta^{++}_L}}
\end{equation}
Here $M_{\Delta^{++}_{L,R}}$ are the masses of $\Delta^{++}_{L,R}$ scalars. The contribution from the heavy neutrino and $W^-_R$ exchange (first Feynman diagram in figure \ref{fig01}) can be written as 
\begin{equation}
A_{NRR} \propto G^2_F \left ( \frac{M_{W_L}}{M_{W_R}} \right )^4 \sum_i \frac{V^{*2}_{ei}}{M_i} 
\end{equation}
The contribution from $N-W_L$ exchange shown by the first diagram in figure \ref{fig01} is given by 
\begin{equation}
A_{NLL} \propto  G^2_F \sum_i \frac{S^2_{ei}}{M_i} 
\end{equation}
The contribution from $\nu-W_R$ exchange shown by the second diagram in figure \ref{fig01} is given by
\begin{equation}
A_{\nu RR} \propto G^2_F \left ( \frac{M_{W_L}}{M_{W_R}} \right )^4 \sum_i \frac{m_i T^{*2}_{ei}}{p^2}
\end{equation}
The so called $\lambda$ contributions come from the first two diagrams in figure \ref{fig02} and are given respectively by 
\begin{equation}
A_{\nu LR} \propto G^2_F \left ( \frac{M_{W_L}}{M_{W_R}} \right )^2 \sum_i \frac{U_{ei} T^*_{ei}}{p}
\end{equation}
\begin{equation}
A_{N LR} \propto G^2_F \left ( \frac{M_{W_L}}{M_{W_R}} \right )^2 \sum_i S_{ei} V^{*}_{ei} \frac{p}{M^2_i}
\end{equation}
out of which only the first one dominates whereas the second contribution can be neglected due to $\frac{p}{M^2_i}$ suppression. The $\eta$ diagram (shown in the last diagram of figure \ref{fig02}) contribution is given by 
\begin{equation}
A_{\eta} \propto G^2_F \tan{\xi} \sum_i \frac{U_{ei} T^*_{ei}}{p}
\end{equation}
where the $W_L-W_R$ mixing parameter $\xi$ is given by
\begin{equation}
\tan{2\xi} = \frac{2v_1 v_2}{v^2_R-v^2_L}
\end{equation}
which is constrained to be $\xi \leq 7.7 \times 10^{-4}$ \cite{rhnlhc} as mentioned above. Here $v_{1,2}, v_{L,R}$ are the vev's of the neutral components of the scalar bidoublet and scalar triplets mentioned in subsection \ref{sec2}. Using the expression for Dirac neutrino mass matrix for LRSM in terms of light neutrino and heavy neutrino mass matrices $M_{\nu}, M_{RR}$ given in \cite{mLRgoran}
\begin{equation}
m_{LR} = M_{RR} \left ( \gamma \frac{M^2_W}{v^2_R}-M^{-1}_{RR} M_{\nu} \right )^{1/2}
\label{eqmLR}
\end{equation}
one can write down all the above expressions in terms of $M_{\nu}, M_{RR}, \gamma$. Combining all the contributions, one can write down the half-life of neutrinoless double beta decay as
\begin{align}
\frac{1}{T^{0\nu}_{1/2}} &= G^{0\nu}_{01} \bigg ( \lvert \mathcal{M}^{0\nu}_\nu  (\eta^L_{\nu} + \eta_{\Delta_L}+\eta^R_{\nu})+\mathcal{M}^{0\nu}_N \eta^L_N \rvert^2 + \lvert \mathcal{M}^{0\nu}_N (\eta^R_N+\eta_{\Delta_R}) \rvert^2  \nonumber \\
& + \lvert \mathcal{M}^{0\nu}_{\lambda} \eta_{\lambda} + \mathcal{M}^{0\nu}_{\eta} \eta_{\eta} \rvert^2 \bigg )
\label{eq:halflife}
\end{align}
where 
$$ \eta^L_{\nu} = \sum_i \frac{m_i U^2_{ei}}{m_e} , \;\;\;\; \eta_{\Delta_L}=  \frac{(M_{\nu}^{II})_{ee}}{M^2_{\Delta^{++}_L}} m_p $$
$$ \eta^R_{\nu}= \left ( \frac{M_{W_L}}{M_{W_R}} \right )^4 \sum_i \frac{m_i T^{*2}_{ei}}{m_e}, \;\;\;\; \eta^L_N=m_p  \sum_i \frac{S^2_{ei}}{M_i} $$
$$ \eta^R_N= m_p\left ( \frac{M_{W_L}}{M_{W_R}} \right )^4 \sum_i \frac{V^{*2}_{ei}}{M_i}, \;\;\;\; \eta_{\Delta_R}= m_p\left ( \frac{M_{W_L}}{M_{W_R}} \right )^4 \sum_i \frac{V^2_{ei} M_i}{M^2_{\Delta^{++}_R}} $$
$$ \eta_{\lambda}=\left ( \frac{M_{W_L}}{M_{W_R}} \right )^2 \sum_i U_{ei} T^*_{ei}, \;\;\;\; \eta_{\eta} =\tan{\xi} \sum_i U_{ei} T^*_{ei}$$
Here $m_e, m_p$ are masses of electron and proton respectively. Also, the nuclear matrix elements involved are denoted by $\mathcal{M}$ the numerical values of which are shown in table \ref{tableNME}. The numerical values of the phase space factor $G^{0\nu}_{01}$ are also shown in the table \ref{tableNME} for different nuclei. In the above equation \eqref{eq:halflife}, the contributions $\eta_{\Delta_L}, \eta^R_N, \eta_{\Delta_R}$ are directly related to the type II seesaw term which also decides the right handed neutrino mass matrix, as seen from equation \eqref{type2}. The contribution $\eta^L_{\nu}$ is the effective light neutrino contributions which acquires mass from both type I and type II seesaw. The remaining contributions arise from the mixing between heavy and light neutrinos through type I seesaw term.
\begin{center}
\begin{table}[htb]
\begin{tabular}{|c|c|c|c|c|c|}
\hline
Isotope & $G^{0\nu}_{01} \; (\text{yr}^{-1})$  &  $\mathcal{M}^{0\nu}_\nu$ & $\mathcal{M}^{0\nu}_N$ & $\mathcal{M}^{0\nu}_{\lambda}$ & $\mathcal{M}^{0\nu}_{\eta}$ \\
\hline
$ \text{Ge}-76$ & $5.77\times10^{-15}$ & $2.58-6.64$ & $233-412$ &$1.75-3.76$ & $235-637$ \\
$ \text{Xe}-136$ & $3.56 \times 10^{-14}$ & $1.57-3.85$ & $164--172$ & $1.92-2.49$ & $370-419$ \\
\hline
\end{tabular}
\caption{Values of phase space factor and nuclear matrix elements used in the analysis.}
\label{tableNME}
\end{table}
\end{center}

Our goal in this work is to point out the new physics contribution to $0\nu \beta \beta$ when type I and type II seesaw both can be equally dominating. This can be very different from the type I or type II dominance cases discussed in earlier works, for example \cite{ndbd1}. Depending on the seesaw mechanism at work, these new physics sources can have different contributions to the neutrinoless double beta decay. It should be noted that the present experimental constrains on the $0\nu \beta \beta$ half-life from the GERDA experiment \cite{GERDA} is
\begin{equation}
T^{0\nu}_{1/2} (\text{Ge}76) > 3.0 \times 10^{25} \; \text{yr}
\end{equation}
Similar bound from the KamLAND-Zen experiment \cite{kamland} is 
\begin{equation}
T^{0\nu}_{1/2} (\text{Xe}136) > 3.4 \times 10^{25} \; \text{yr}
\label{ndbdbound}
\end{equation}
More recently, KamLAND-Zen collaboration has updated their earlier estimates with an improved lower limit on $0\nu \beta \beta$ half-life \cite{kamland2}
$$ T^{0\nu}_{1/2} (\text{Xe}136) > 1.1 \times 10^{26} \; \text{yr} $$

\subsection{Charged Lepton flavour Violation in MLRSM}
\label{sec:lfvlhc}
Lepton flavour violation (LFV) in MLRSM were studied in details in previous works including \cite{LFVLR}. Within this model, there are several possible LFV processes like $\mu \rightarrow e\gamma, \mu \rightarrow 3e$. Here we consider $\mu \rightarrow 3e$ process mediated by doubly charged bosons in MLRSM. The current experimental bound on this process from SINDRUM collaboration \cite{sindrum} is
\begin{equation}
 \text{BR}(\mu \rightarrow 3e) < 10 \times 10^{-12}
 \label{meeebound}
 \end{equation}
The branching ratio for the $\mu \rightarrow 3e$ process induced by doubly charged bosons $\Delta^{++}_L, \Delta^{++}_R$ is given by \cite{LFVLR}
\begin{equation}
\text{BR}(\mu \rightarrow 3e) = \frac{1}{2} \lvert h_{\mu e} h^{*}_{ee} \rvert^2 \left ( \frac{M^4_{W_L}}{M^4_{\Delta^{++}_L}}+\frac{M^4_{W_L}}{M^4_{\Delta^{++}_R}} \right )
\label{eqBR}
\end{equation}
where the couplings $h$ are given by
\begin{equation}
h_{ij} = \sum_n \left ( V \right )_{ni} \left ( V \right )_{nj} \sqrt{\left(\frac{M_i}{M_{W_R}}\right)^2}
\label{eqhij}
\end{equation}
In equation \eqref{eqBR}, $M_{\Delta^{++}_{L,R}}$ are the masses of $\Delta^{++}_{L,R}$ and in equation \eqref{eqhij}, $V$ is the mixing matrix and $M_i$ are right handed neutrino masses defined in the previous section. In a previous work \cite{ndbd00}, the experimental bound on this LFV process was incorporated to restrict $M^{\text{heaviest}}_i/M_{\Delta}$, where $\frac{1}{M^2_{\Delta}} = \frac{1}{M^2_{\Delta^{++}_L}}+\frac{1}{M^2_{\Delta^{++}_R}}$. It was found that for most of the parameter space, $M^{\text{heaviest}}_i/M_{\Delta}<0.1$ with $M_{W_R} = 3.5$ TeV. Assuming $M_{\Delta^{++}_L} = M_{\Delta^{++}_R} = M_{\delta}$, the above bound will become $M^{\text{heaviest}}_i/M_{\delta}<0.1/\sqrt{2}$. However, this bound was calculated only with the assumption that $U_R = U_L$ and hence may not be applicable in a general case where both type I and type II seesaw terms contribute to light neutrino masses. Similarly, the branching ratio for $\mu \rightarrow e\gamma$ is given by \cite{ndbd10}
\begin{equation}
\text{BR}(\mu \rightarrow e\gamma) =\frac{3\alpha_{\text{em}}}{2\pi} \left ( \lvert G^{\gamma}_L \rvert^2+ \lvert G^{\gamma}_R \rvert^2 \right )
\label{eqBR1}
\end{equation}
where $\alpha_{\text{em}}=e^2/4\pi$ and the form factors $G^{\gamma}_{L,R}$ are given by
\begin{equation}
G^{\gamma}_L = \sum^3_{i=1}  \left ( S^*_{\mu i}S_{ei} G^{\gamma}_1 (x_i)-V_{\mu i}S_{e i} \xi e^{i\zeta} G^{\gamma}_2 (x_i)\frac{M_i}{m_{\mu}} + (V)_{\mu i}(V)^*_{ei} y_i \bigg[\frac{2}{3} \frac{M^2_{W_L}}{M^2_{\Delta^{++}_L}}+\frac{1}{12} \frac{M^2_{W_L}}{M^2_{\Delta^{+}_L}} \bigg] \right )
\nonumber
\end{equation}
\begin{align}
G^{\gamma}_R &= \sum^3_{i=1}   \bigg ( (V)_{\mu i}(V)^*_{ei} \lvert \xi^2 \rvert G^{\gamma}_1 (x_i)-S^*_{\mu i}(V)^*_{e i} \xi e^{-i\zeta} G^{\gamma}_2 (x_i)\frac{M_i}{m_{\mu}}  \nonumber \\
&+ (V)_{\mu i}(V)^*_{ei} \bigg[\frac{M^2_{W_L}}{M^2_{W_R}}G^{\gamma}_1 (y_i)+\frac{2y_i}{3} \frac{M^2_{W_L}}{M^2_{\Delta^{++}_R}} \bigg]  \bigg)
\end{align}
In the above expressions, $x_i \equiv (M_i/M_{W_L})^2$, $y_i\equiv (M_i/M_{W_R})^2$, $\zeta$ is the phase of vev $v_2$ (taken to be zero here), $m_{\mu}$ is the muon mass, $S$ is the light-heavy neutrino mixing matrix and $\xi$ is the $W_L-W_R$ mixing parameter defined earlier. In the earlier works, the elements of $S$ and the mixing $\xi$ were assumed to be negligible. But here we consider them in the analysis of LFV similar to the way there were included in the $0\nu \beta \beta$ amplitudes. The loop functions $G^{\gamma}_{1,2}$ are given by
\begin{equation}
G^{\gamma}_1 (a) = -\frac{2a^3+5a^2-a}{4(1-a)^3}-\frac{3a^3}{2(1-a)^4} \ln{a}
\nonumber
\end{equation}
\begin{equation}
G^{\gamma}_2 (a) = \frac{a^2-11a+4}{2(1-a)^2}-\frac{3a^2}{(1-a)^3}\ln{a}
\end{equation} 
The experimental bound on this LFV process from MEG collaboration \cite{MEG} is 
\begin{equation}
\text{BR}(\mu \rightarrow e\gamma) < 5.7 \times 10^{-13} 
\label{megbound}
\end{equation} 
This upper bound is slightly improved in the latest estimate by MEG collaboration to $4.2 \times 10^{-13}$ \cite{MEG2}.

\subsection{Collider Constraints}
\label{sec:collider}
Apart from LFV bounds on the ratio $M^{\text{heaviest}}_i/M_{\Delta}$, there exists other experimental bounds on the new particles of LRSM. The most stringent bound on the additional charged vector boson $W_R$ comes from the $K-\bar{K}$ mixing: $M_{W_R} > 2.5$ TeV \cite{kkbar}. Direct searches at LHC also put similar constraints on the mass of $W_R$ boson. Dijet resonance search by ATLAS puts a bound $M_{W_R} > 2.45$ TeV at $95\%$ CL \cite{dijetATLAS}. This bound can however be relaxed to $M_{W_R} \geq 2$ TeV if $g_R \approx 0.6 g_L$. There are other bounds on $M_{W_R}$ coming from other searches in LHC experiments, but they are weaker than the dijet resonance bound. For example, the CMS experiment at the LHC excludes some parameter space in the $M^{\text{lightest}}_i-M_{W_R}$ plane from the search of $pp \rightarrow l^{\pm} l^{\pm} j j$ processes mediated by heavy right handed neutrinos at 8 TeV centre of mass energy \cite{CMSNRWR}. Similarly, the doubly charged scalars also face limits from CMS and ATLAS experiments at LHC:
$$ M_{\Delta^{\pm \pm}} \geq 445 \; \text{GeV} \; (409 \; \text{GeV}) \; \text{for} \; \text{CMS (ATLAS)} $$
These limits have been put by assuming $100\%$ leptonic branching factions \cite{hdlhc}.

 A review of heavy neutrino searches at colliders both in the presence and absence of additional gauge interactions can be found in \cite{futureCollider}. As discussed in \cite{futureCollider}, direct searches for $W_L-\nu_R$ mediated same-sign dilepton plus dijet at the LHC with 8 TeV centre of mass energy can constrain the heavy neutrino mixing with muon type light neutrino to be less than $10^{-2}-\mathcal{O}(1)$ for heavy neutrino masses from 30 GeV to 500 GeV. The bounds are slightly weaker for the mixing parameter of electron type neutrino with the heavy neutrinos. For smaller heavy-light neutrino mixing, the production cross section for such a process can be enhanced in the presence of additional gauge interactions, like in the MLRSM discussed above. The heavy right handed neutrinos with $SU(2)_R$ gauge interactions are constrained by direct searches at LHC. For example, the search for $W_R \rightarrow l_R \nu_R$ at ATLAS and CMS constrains the right handed neutrino masses to be around 1 TeV \cite{rhnlhc}. In fact, right handed neutrino mass as high as 1.8 TeV can be excluded by 8 TeV LHC data. However, such bounds are valid for specific $W_R$ masses as can be seen from the exclusion plots in $M^{\text{lightest}}_i-M_{W_R}$ plane given in \cite{CMSNRWR}. As discussed in \cite{futureCollider}, the LHC at 14 TeV centre of mass energy should be able to prove heavy neutrino masses upto around 3 TeV along with $W_R$ boson mass upto 5 TeV. At this point, it is worth noting that the lower bounds on the scalar masses (apart from SM Higgs and $\delta^0_R $) could be more severe from perturbativity bounds than the direct search bounds, specially with TeV scale $W_R$ \cite{perturbLRSM}.

\section{Combination of Type I and Type II Seesaw}
\label{sec:type12}
As mentioned above, almost all the earlier works discussing $0\nu \beta \beta$ and LFV within MLRSM have considered either type I or type II seesaw dominance at a time. However, the new physics contribution to $0\nu \beta \beta$ can be very different from these two simplest scenarios if type I and type II seesaw contributions to light neutrino masses are comparable. In this case, one can not relate the diagonalising matrices of light and heavy neutrino mass matrices. Some simple relations relating different mass matrices involved in the formula for light neutrino masses in MLRSM given by equation \eqref{type2} were discussed in \cite{mLRgoran}. One useful parametrisation of the Dirac neutrino mass matrix in the presence of type I+II seesaw was studied by the authors of \cite{akhwer}. In another work \cite{akhfri}, relations between type I and type II seesaw mass matrices were derived by considering the Dirac neutrino mass matrix to be known. If the Dirac neutrino mass matrix $m_{LR}$ is not known, then we can still choose at least one of the type I and type II seesaw mass matrices arbitrarily due to the freedom we have in choosing $m_{LR}$ that appears in the type I seesaw term. After choosing one the seesaw mass matrices, the other gets completely fixed if the light neutrino mass matrix is completely known. Interestingly in MLRSM, once we choose the type II seesaw mass matrix, we can calculate $M_{RR}$ using its relation between type II seesaw mass matrix \eqref{type2} and from that $M_{RR}$, the Dirac neutrino mass matrix $m_{LR}$ can be derived using \eqref{eqmLR}.

The Pontecorvo-Maki-Nakagawa-Sakata (PMNS) leptonic mixing matrix is related to the diagonalising 
matrices of neutrino and charged lepton mass matrices $U_{\nu}, U_l$ respectively, as
\begin{equation}
U_{\text{PMNS}} = U^{\dagger}_l U_{\nu}
\label{pmns0}
\end{equation}
The PMNS mixing matrix can be parametrised as
\begin{equation}
U_{\text{PMNS}}=\left(\begin{array}{ccc}
c_{12}c_{13}& s_{12}c_{13}& s_{13}e^{-i\delta}\\
-s_{12}c_{23}-c_{12}s_{23}s_{13}e^{i\delta}& c_{12}c_{23}-s_{12}s_{23}s_{13}e^{i\delta} & s_{23}c_{13} \\
s_{12}s_{23}-c_{12}c_{23}s_{13}e^{i\delta} & -c_{12}s_{23}-s_{12}c_{23}s_{13}e^{i\delta}& c_{23}c_{13}
\end{array}\right) U_{\text{Maj}}
\label{matrixPMNS}
\end{equation}
where $c_{ij} = \cos{\theta_{ij}}, \; s_{ij} = \sin{\theta_{ij}}$ and $\delta$ is the leptonic Dirac CP phase. The diagonal matrix $U_{\text{Maj}}=\text{diag}(1, e^{i\alpha}, e^{i(\beta+\delta)})$  contains the Majorana CP phases $\alpha, \beta$ which remain undetermined at neutrino oscillation experiments. For diagonal charged lepton mixing matrix, the neutrino mass diagonalisation matrix can be identified with the leptonic mixing matrix $U_{\text{PMNS}} = U_{\nu}$. In that case, the light neutrino mass matrix can be constructed as
\begin{equation}
M_{\nu}=U_{\text{PMNS}}M^{\text{diag}}_{\nu} U^T_{\text{PMNS}}
\end{equation}
where $M^{\text{diag}}_{\nu} = \text{diag}(m_1, m_2, m_3)$ is the diagonal light neutrino mass matrix. It should be noted that, here we are ignoring the non-unitary effects due to heavy-light neutrino mixing and using the parametric form $U_{\text{PMNS}}$ as the diagonalising matrix of light neutrino mass matrix. The actual light neutrino mixing matrix $U$ is non-unitary due to the presence of heavy-light neutrino mixing, and related to $U_L=U_{\text{PMNS}}$ through \eqref{mixingmatrix}.

If the type II seesaw mass matrix gives rise to a mixing matrix $U_{II}$, then we can write down the type II seesaw mass matrix as $M^{II}_{\nu}=U_{II} M^{II(\text{diag})}_{\nu} U^T_{II}$ where $M^{II(\text{diag})}_{\nu} = X M^{\text{diag}}_{\nu}$. Here $X$ is a numerical factor which decides the strength of type II seesaw contribution to light neutrino masses. In MLRSM, the type II seesaw mass matrix is proportional to the right handed Majorana neutrino mass matrix
$$ \gamma (M_{W_L}/v_{R})^{2}M_{RR} = M^{II}_{\nu}$$
as seen from equation \eqref{type2}. We consider a general diagonalising matrix $U_{II}$ for $3\times 3$ right handed neutrino mass matrix $M_{RR}$. This diagonalising matrix $U_{II}$ can be parametrised in a way similar to the PMNS mixing matrix shown above. The matrix $U_{II}$ can have arbitrary angles and phases, unobserved in light neutrino oscillations. For simplicity, we parametrise it with three angles $\phi_{12}, \phi_{23}, \phi_{13}$ only. Once the structure of type II seesaw mass matrix is chosen, the type I seesaw mass matrix automatically gets fixed by the requirement that their combination should give rise to the correct light neutrino mass matrix. The eigenvalues of the right handed neutrino mass matrix can be written as
\begin{equation}
\text{diag}(M_1, M_2, M_3) = \frac{1}{\gamma} \left(\frac{v_R}{M_{W_L}}\right )^{2} X M^{\text{diag}}_{\nu}
\end{equation}
For normal hierarchy, the diagonal mass matrix of the light neutrinos can be written  as $M^{\text{diag}}_{\nu} 
= \text{diag}(m_1, \sqrt{m^2_1+\Delta m_{21}^2}, \sqrt{m_1^2+\Delta m_{31}^2})$ whereas for inverted hierarchy 
 it can be written as $M^{\text{diag}}_{\nu} = \text{diag}(\sqrt{m_3^2+\Delta m_{23}^2-\Delta m_{21}^2}, 
\sqrt{m_3^2+\Delta m_{23}^2}, m_3)$. The mass squared differences can be taken from the global fit neutrino oscillation data shown in table \ref{tab:data1} shown above, leaving the lightest neutrino mass as free parameter in $M^{\text{diag}}$. Thus, the right handed neutrino mass matrix can be written in terms of five free parameters: the lightest neutrino mass, three angles $\phi_{12}, \phi_{23}, \phi_{13}$ and the seesaw relative strength factor $X \frac{1}{\gamma} \left(\frac{v_R}{M_{W_L}}\right )^{2}$. 

\begin{table}[!h]
\centering
\begin{tabular}{|c|c|c|}
\hline
Parameters &  Values (NH) & Values (IH) \\
\hline
$\frac{\Delta m^2_{21}}{10^{-5} {\rm eV}^2}$ & 7.60  & 7.60\\
$\frac{\lvert \Delta m^2_{31}\rvert}{10^{-3} {\rm eV}^2} $ & 2.48 & 2.38 \\
$\sin^2 \theta_{12}$ & 0.323  & 0.323\\
$\sin^2 \theta_{23}$ & 0.567 & 0.573 \\
$\sin^2 \theta_{13}$ & 0.0234 & 0.024 \\
$p$ & 100 MeV & 100 MeV \\ 
$M_{W_L}$ & 80.4 GeV & 80.4 GeV \\
$M_{W_R}$ &  3.5 TeV & 3.5 TeV \\
\hline
\end{tabular}
\caption{Numerical values of several parameters used in the calculation of $m^{\text{eff}}$ for $0\nu \beta \beta$.}
\label{param}
\end{table}

\section{Numerical Analysis}
\label{sec3}
In the present work, we consider equal dominance of type I and type II seesaw contribution to light neutrino masses. The analysis of $0\nu \beta \beta$ and LFV for individual seesaw dominance can be found in several earlier works. As discussed in the previous section, we first choose the type II seesaw mass matrix $M^{II}_{\nu}=U_{II} M^{II(\text{diag})}_{\nu} U^T_{II}$ where $M^{II(\text{diag})}_{\nu} = X M^{\text{diag}}_{\nu}$. Assuming $U_{II}$ to be an orthogonal matrix, the parametrisation of $M^{II}_{\nu}$ in this particular way involves five free parameters: three angles in $U_{II}$, lightest neutrino mass and $X$. The right handed neutrino mass matrix $M_{RR}$ can also be constructed with five free parameters as discussed above. Once $M_{RR}$ is constructed like this, we can find the Dirac neutrino mass matrix given given by equation \eqref{eqmLR}. Since this involves both $M_{RR}$ and $M_{\nu}$ one requires three more free parameters: the leptonic CP phases contained in $M_{\nu}$ after using the best fit values of the leptonic mixing angles and mass squared differences. Once $m_{LR}, M_{RR}, M_{\nu}$ are constructed, one can find various mixing matrices $U, V, S, T$ discussed in the previous section in terms of eight free parameters. Fixing the charged triplet scalar and right handed gauge boson masses, we then calculate the amplitudes of $0\nu \beta \beta$ and LFV processes. We repeat the same calculation for different benchmark values of $M_{\Delta}, M_{W_R}$ and show the allowed parameter space after incorporating different experimental constraints. 

Once the scale of left-right symmetry is chosen, one can fix the light and heavy neutrino spectrum by fixing two free parameters: the lightest neutrino mass $m_{\text{lightest}}$ and $X/\gamma$.  The heaviest right handed neutrino mass can be written in terms of the heaviest neutrino mass as
\begin{equation}
M^{\text{heaviest}}_i = \frac{X}{\gamma} \left (\frac{v_R}{M_{W_L}} \right )^2 m^{\text{heaviest}}_i
\end{equation}
Since the right handed neutrino masses are generated through their couplings with $\Delta_R$, the maximum value of the heaviest right handed neutrino can be $M_{\text{heaviest}} \geq \sqrt{4\pi} v_R$. Here $\sqrt{4\pi}$ is the maximum perturbative value of Yukawa coupling involved. Considering the lowest possible value of $M_{\text{heaviest}}$ to be $100$ GeV, we arrive at the following range of allowed values of the factor $X/\gamma$
\begin{equation}
\frac{100}{m^{\text{heaviest}}_i} \left(\frac{M_{W_L}}{v_R}\right)^2 \leq \frac{X}{\gamma} \leq \sqrt{4\pi} \frac{M^2_{W_L}}{v_R m^{\text{heaviest}}_i}
\label{xgamma}
\end{equation}

In the present work, we fix the left-right symmetry scale $v_R$ and other parameters shown in table \ref{param} and then vary the other free parameters in the range shown in table \ref{param2}. Choice of parameters in table \ref{param}, \ref{param2} also fixes the range of $X/\gamma$ given by equation \eqref{xgamma}. We then calculate the $0\nu \beta \beta$ half life as well LFV branching ratios for the entire parameter space. We also constrain the parameter space from the requirement of fulfilling experimental lower bound on $0\nu \beta \beta$ half-life and upper bound on LFV branching ratios. For a comparison with earlier results, we specifically choose the parameter $r=\frac{M^{\text{heaviest}}_i}{M_{\Delta}} \equiv \frac{M_N}{M_{\Delta}}$ and show its allowed range. We further show the allowed range of $X/\gamma$, the factor which decides the strength of type II seesaw term.

\begin{table}[!h]
\centering
\begin{tabular}{|c|c|}
\hline
Parameters &  Range \\
\hline
$M_{\Delta^{++}_{L,R}}$ & 500 GeV - $\sqrt{4\pi}v_R$\\
$M^{\text{heaviest}}_i$ & 100 GeV - $\sqrt{4\pi}v_R$\\
$m_{\text{lightest}}$ & $10^{-6}-10^{-1}$ eV \\
$ \delta, \alpha, \beta $ & $0-2\pi$ \\
$\phi_{ij}$ & $0-\pi/4$ \\
\hline
\end{tabular}
\caption{Range of numerical values of several parameters used in the calculation of $T^{0\nu}_{1/2}$ for $0\nu \beta \beta$ as well as LFV branching ratios.}
\label{param2}
\end{table}

\begin{figure}[!h]
\centering
\begin{tabular}{cc}
\epsfig{file=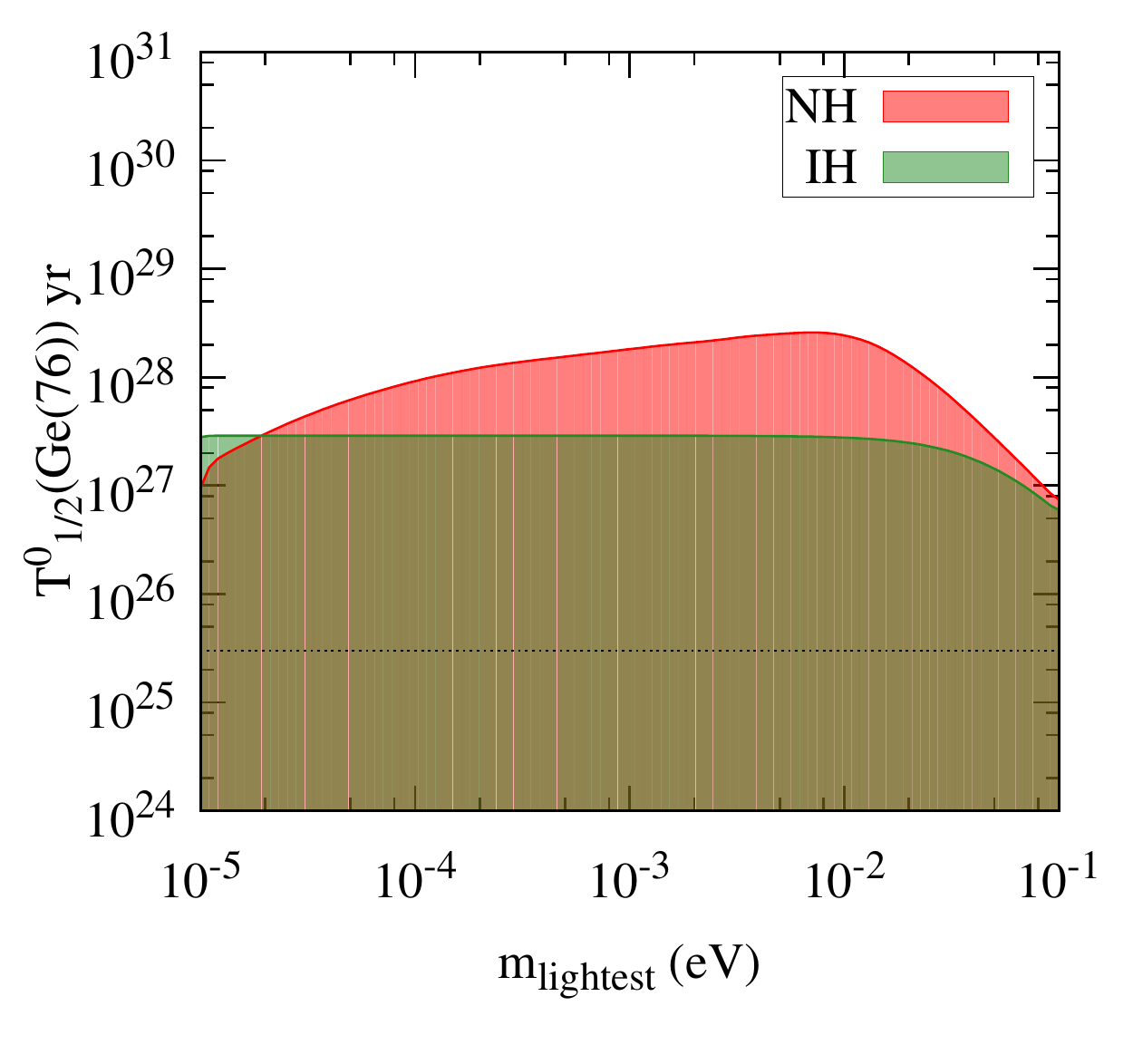,width=0.5\textwidth,clip=}&
\epsfig{file=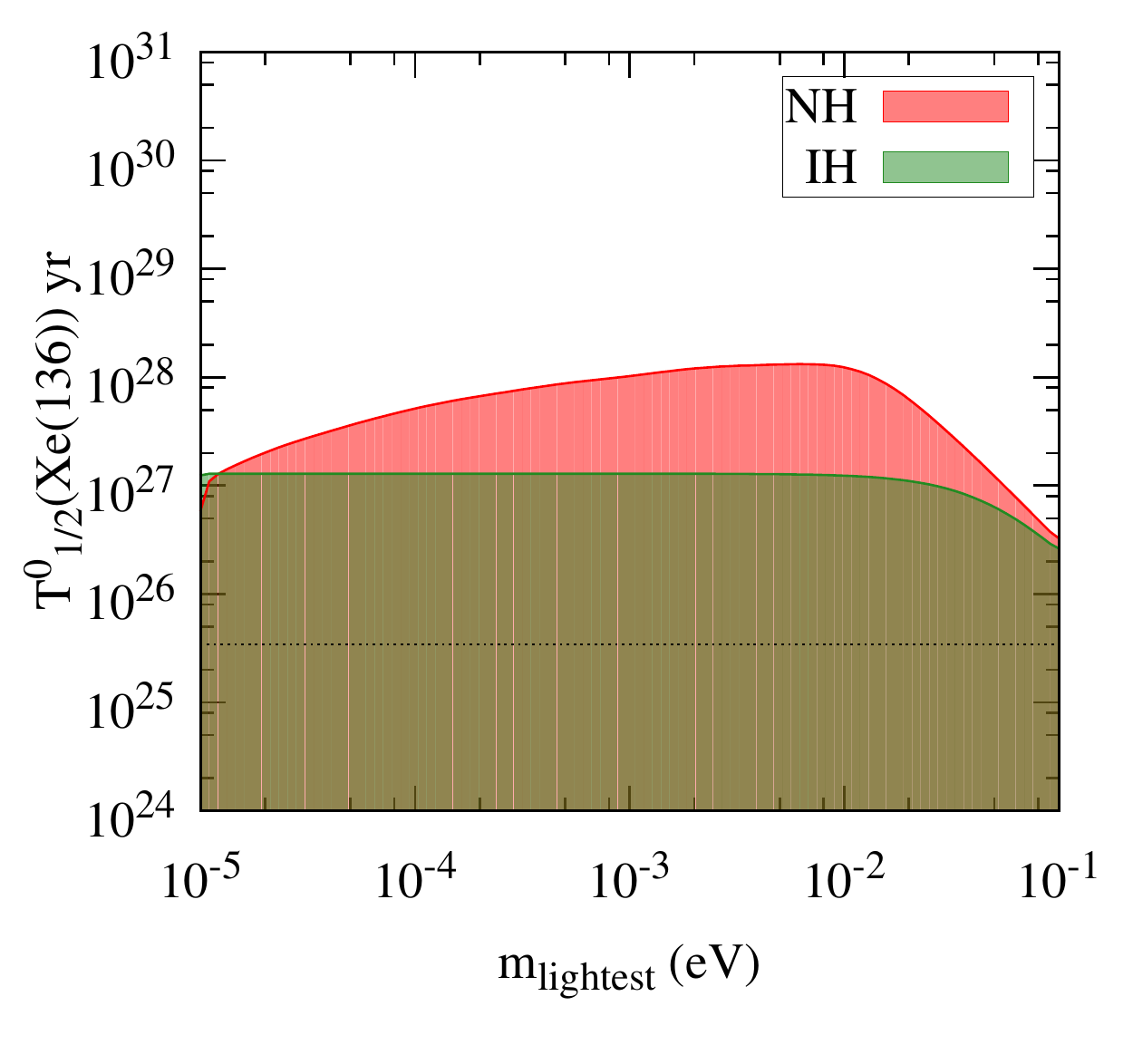,width=0.5\textwidth,clip=} \\
\end{tabular}
\caption{Total contribution to neutrinoless double beta decay half-life with type I+II seesaw. The horizontal lines in the left and right panels of the figure correspond to experimental lower bounds mentioned in \cite{GERDA} and \cite{kamland} respectively.}
\label{fig3}
\end{figure}

\begin{figure}[!h]
\centering
\begin{tabular}{cc}
\epsfig{file=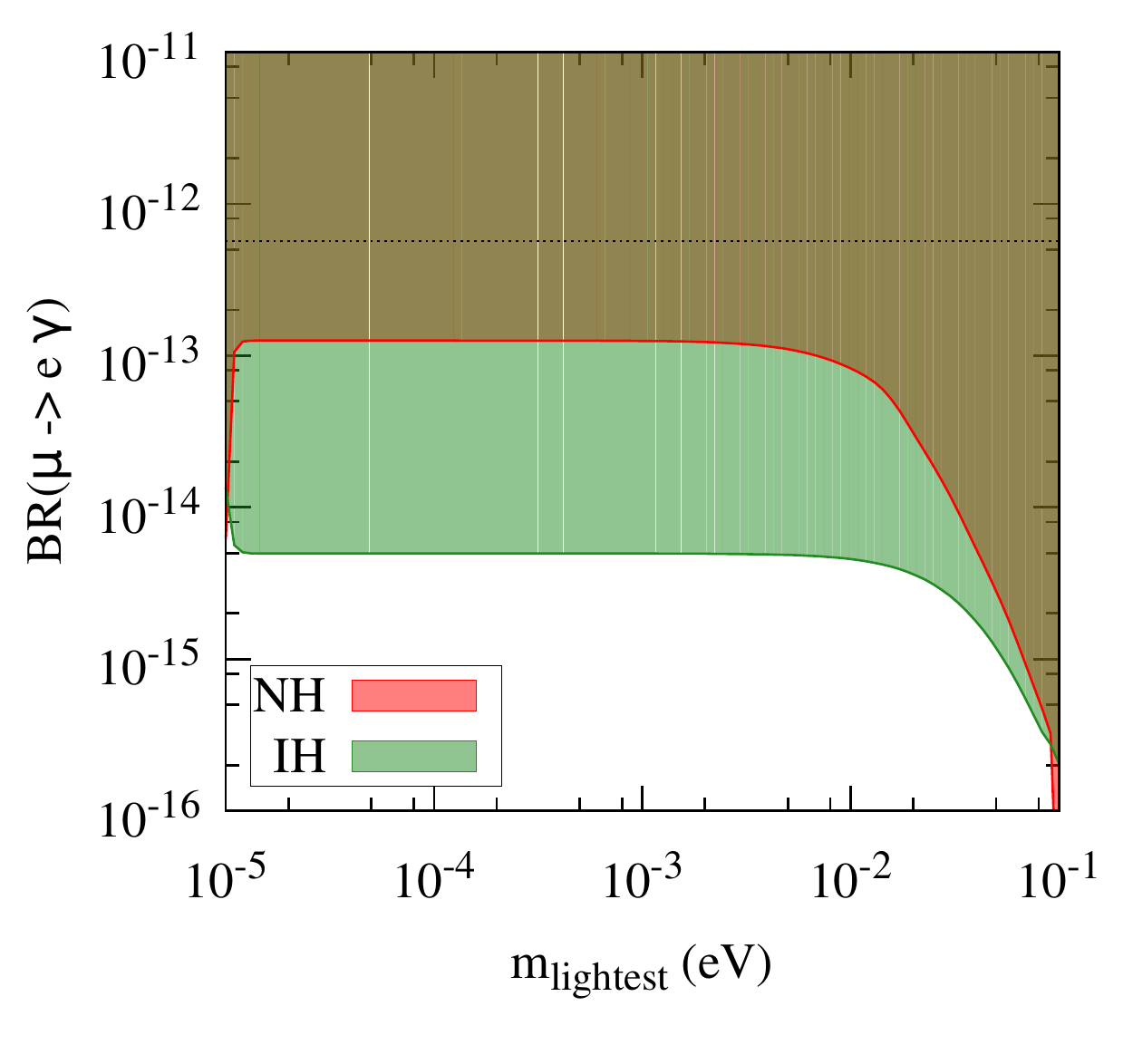,width=0.5\textwidth,clip=}&
\epsfig{file=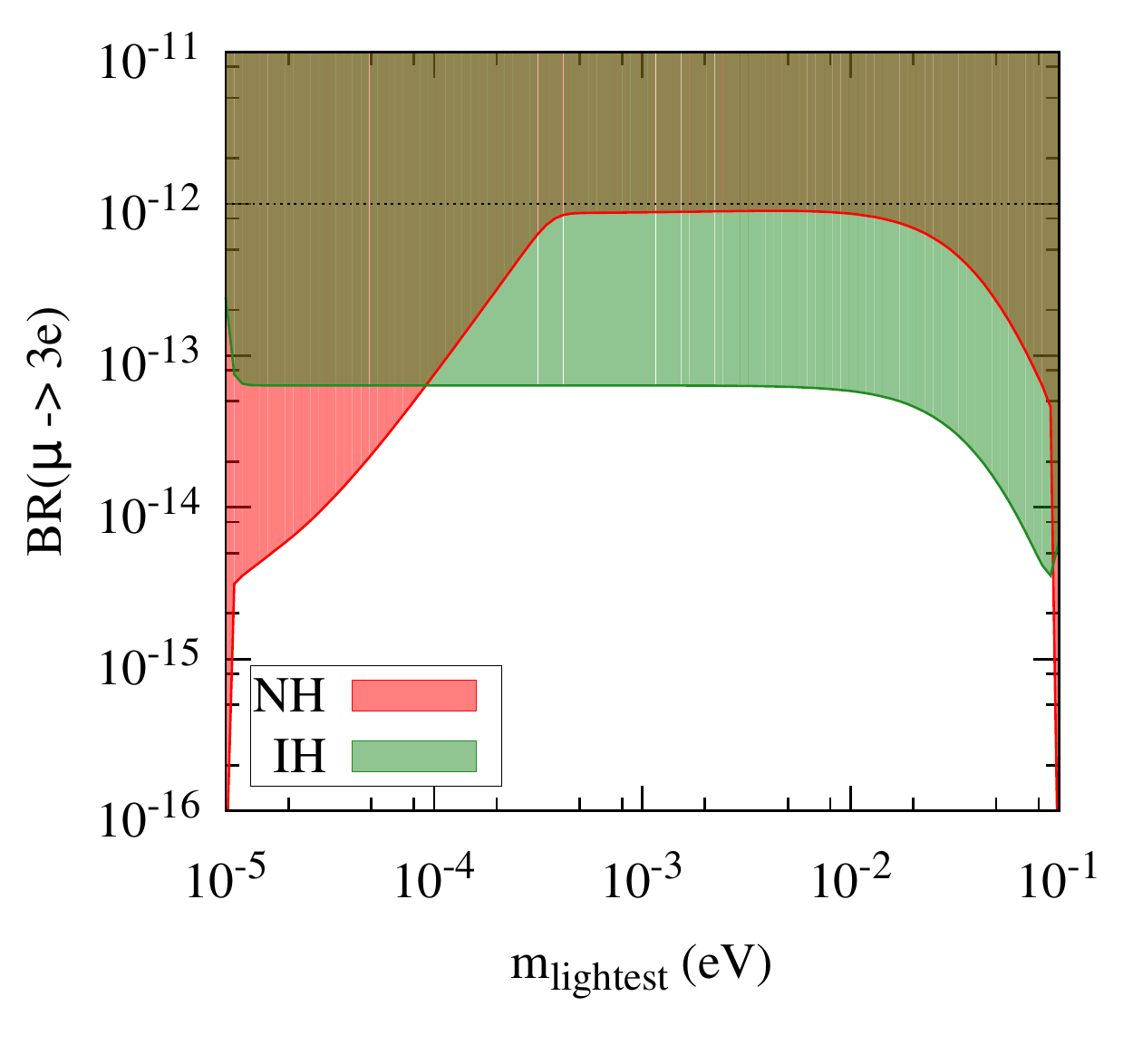,width=0.5\textwidth,clip=} \\
\end{tabular}
\caption{Total contribution to charged lepton flavour violation with type I+II seesaw. The horizontal lines in the left and right panels of the figure correspond to experimental lower bounds mentioned in \cite{sindrum} and \cite{MEG} respectively.}
\label{fig4}
\end{figure}

\begin{figure}[!h]
\centering
\begin{tabular}{cc}
\epsfig{file=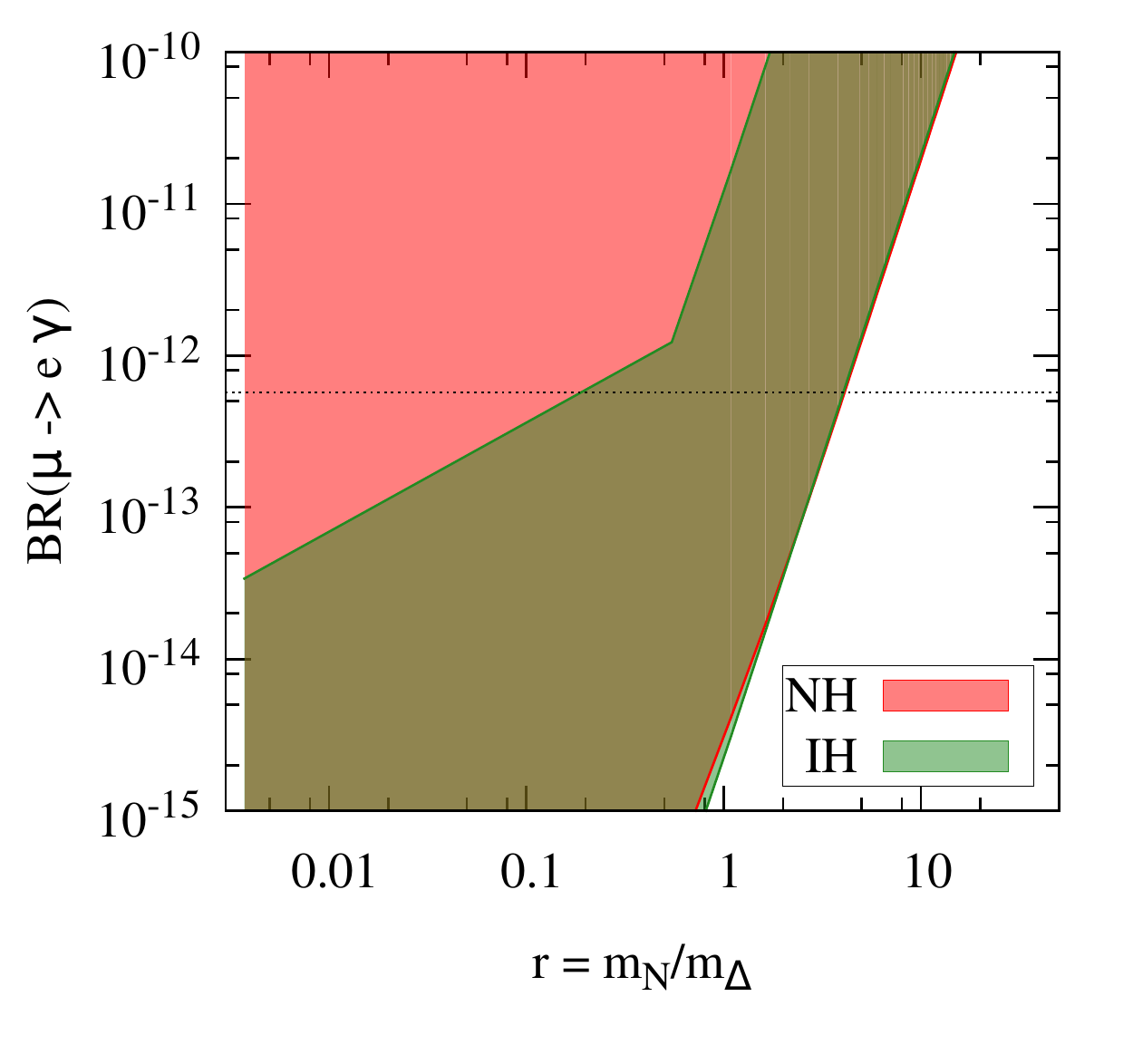,width=0.5\textwidth,clip=}&
\epsfig{file=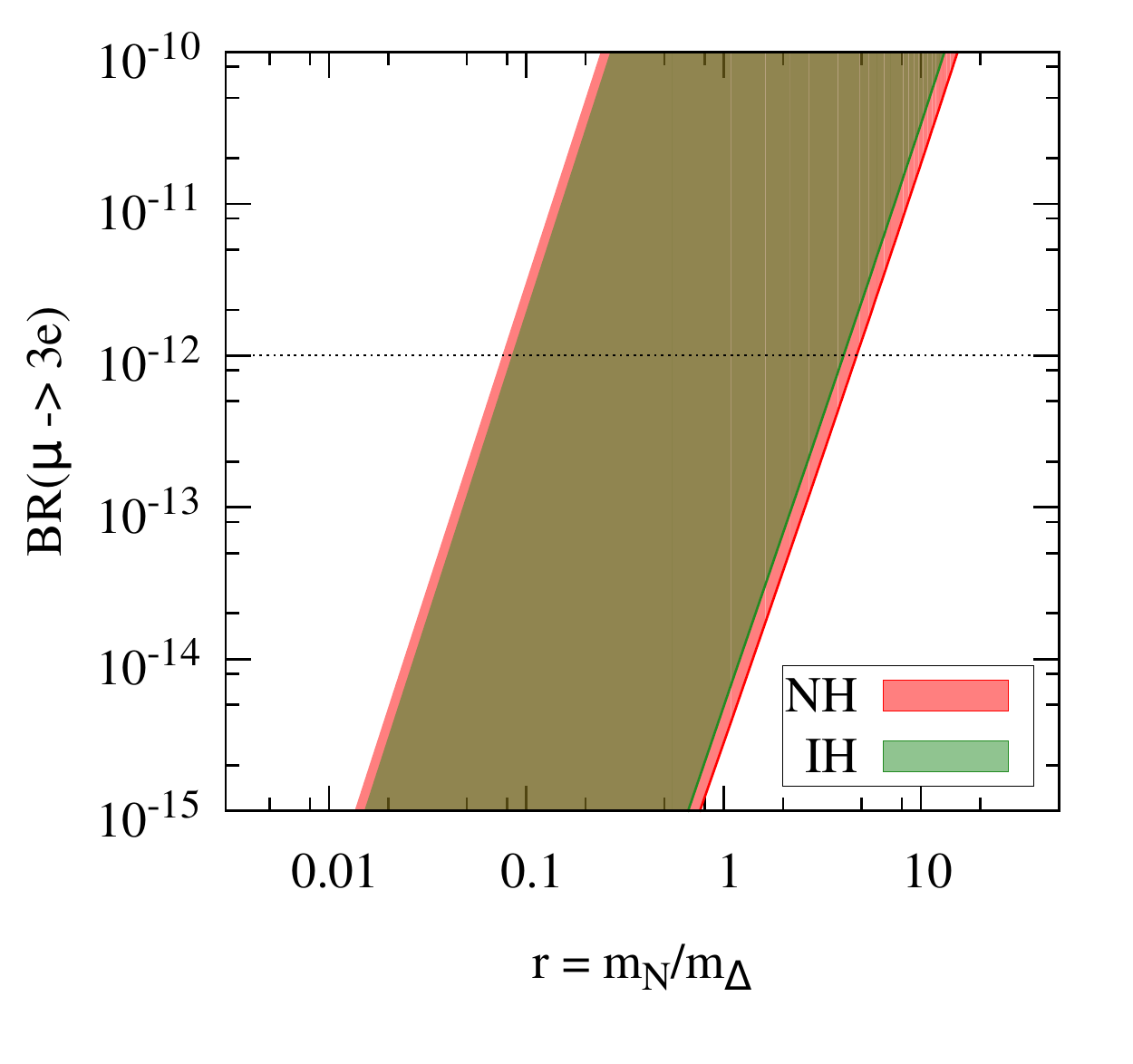,width=0.5\textwidth,clip=} \\
\end{tabular}
\caption{Total contribution to charged lepton flavour violation with type I+II seesaw shown as a function of $r$. The horizontal lines in the left and right panels of the figure correspond to experimental lower bounds mentioned in \cite{sindrum} and \cite{MEG} respectively.}
\label{fig41}
\end{figure}

\begin{figure}[!h]
\centering
\begin{tabular}{cc}
\epsfig{file=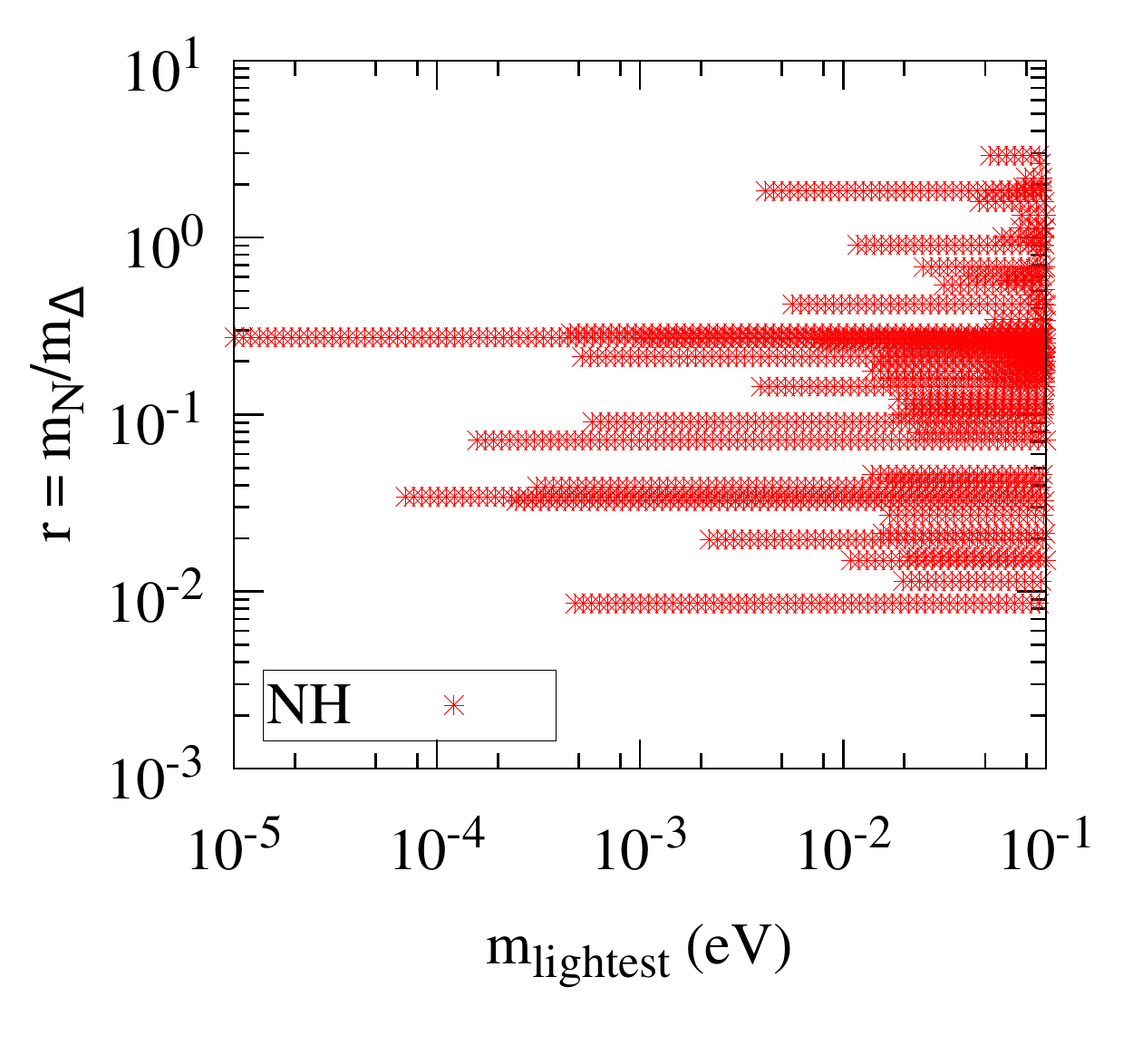,width=0.5\textwidth,clip=}&
\epsfig{file=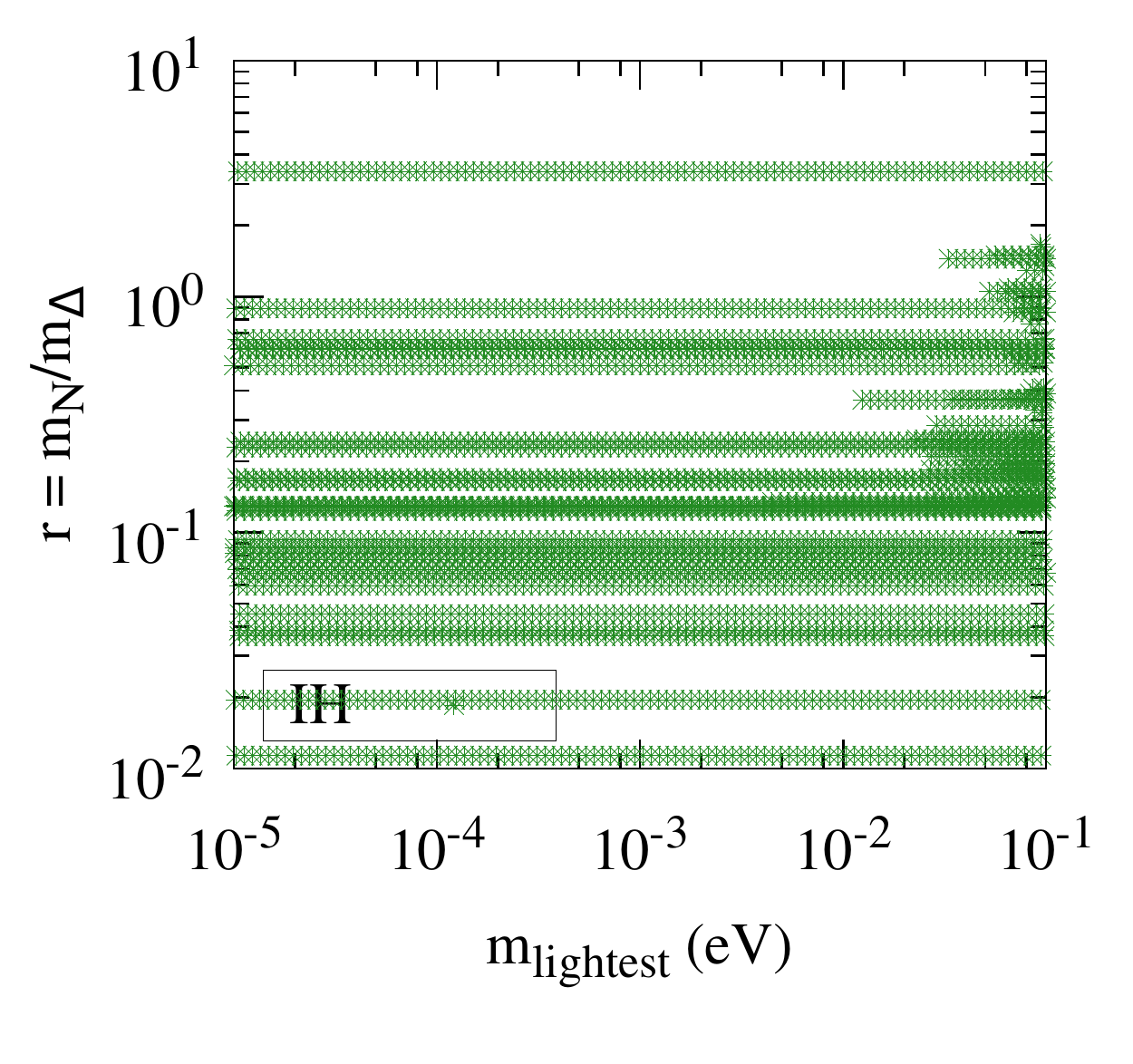,width=0.5\textwidth,clip=} \\
\end{tabular}
\caption{Allowed parameter space in $r-m_{\text{lightest}}$ plane from constraints on neutrinoless double beta decay half-life and charged lepton flavour violation with type I+II seesaw.}
\label{fig5}
\end{figure}

\begin{figure}[!h]
\centering
\begin{tabular}{cc}
\epsfig{file=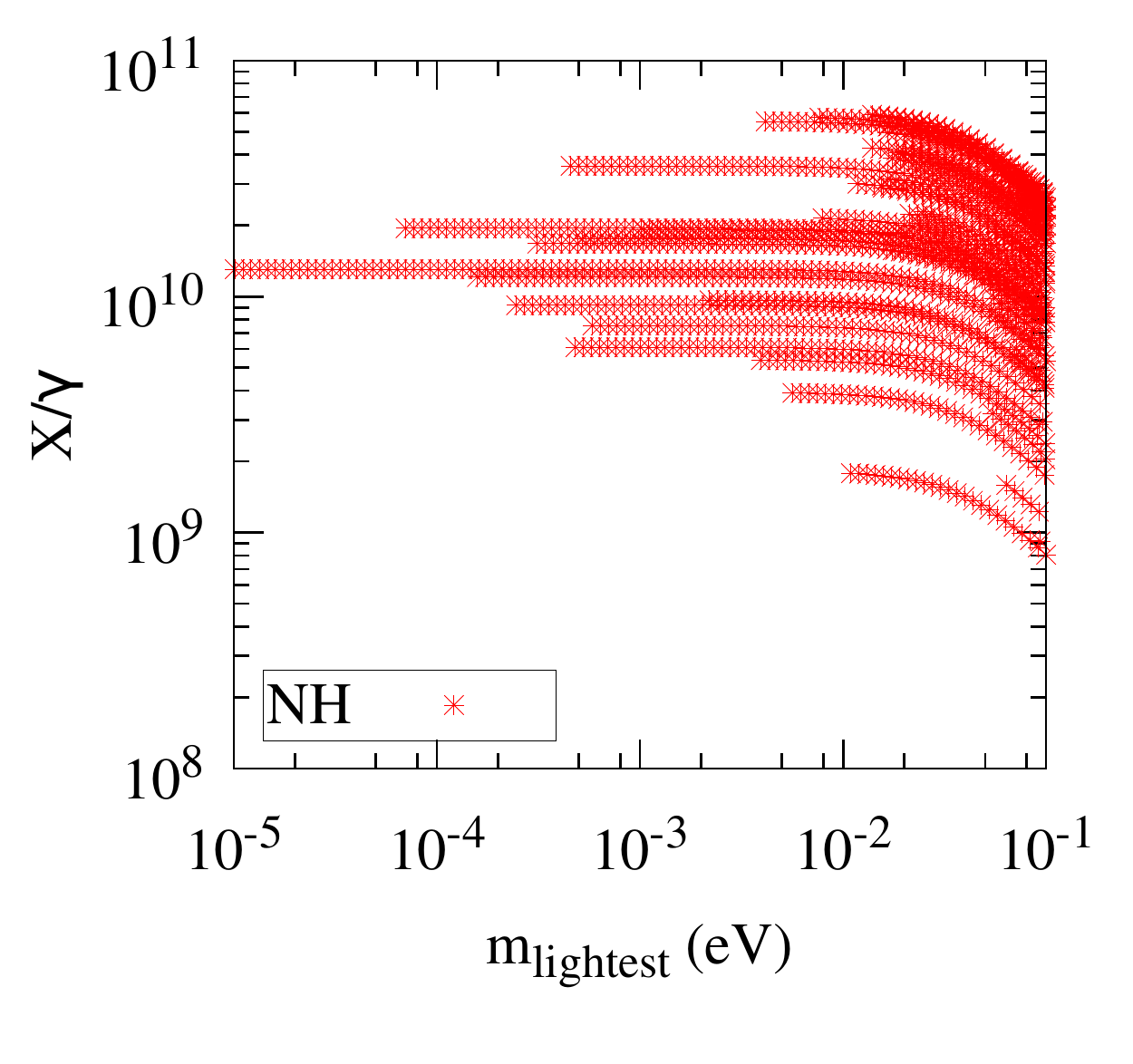,width=0.5\textwidth,clip=}&
\epsfig{file=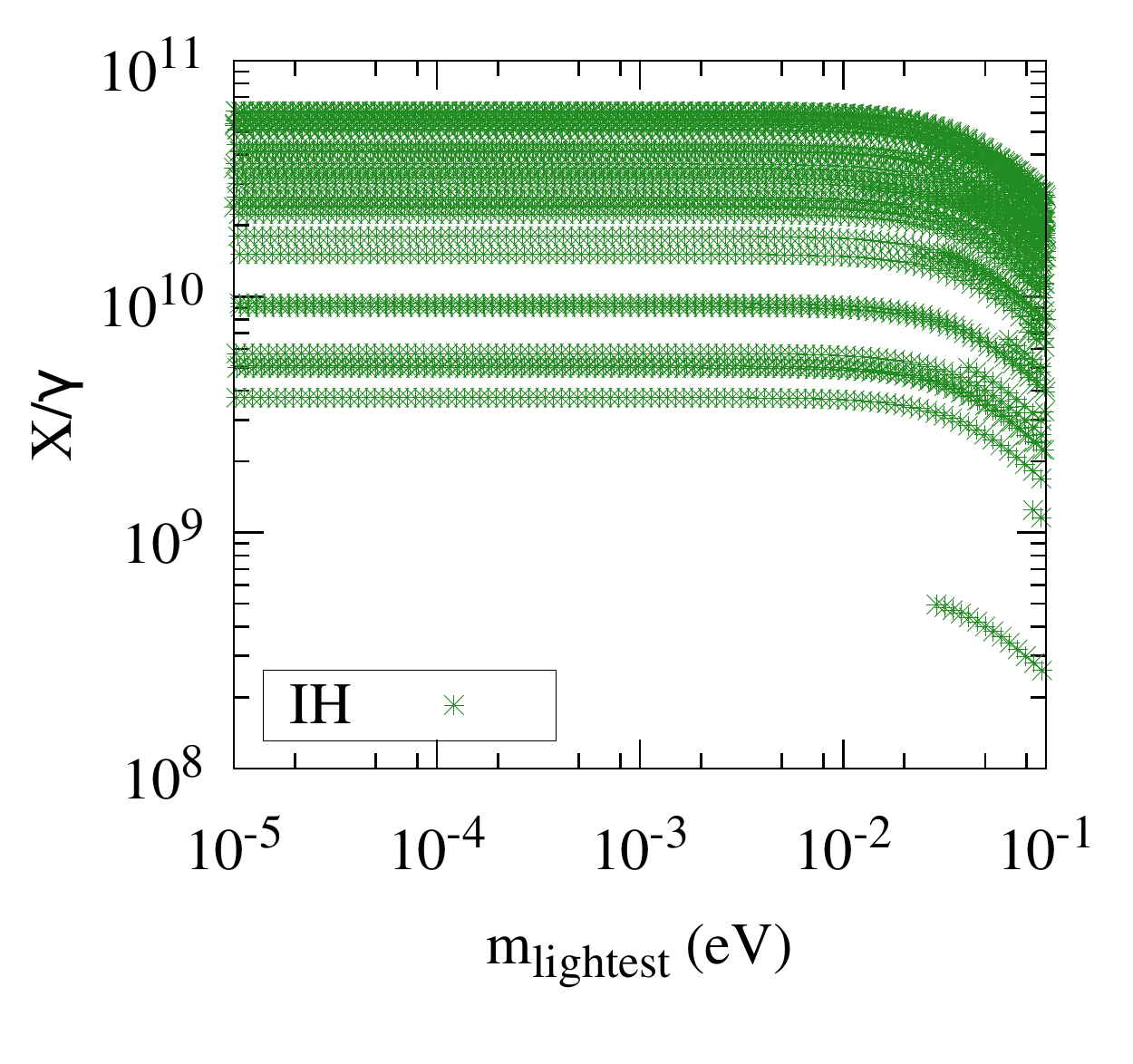,width=0.5\textwidth,clip=} \\
\end{tabular}
\caption{Allowed parameter space in $X/\gamma-m_{\text{lightest}}$ plane from constraints on neutrinoless double beta decay half-life and charged lepton flavour violation with type I+II seesaw.}
\label{fig6}
\end{figure}

\begin{figure}[!h]
\centering
\begin{tabular}{cc}
\epsfig{file=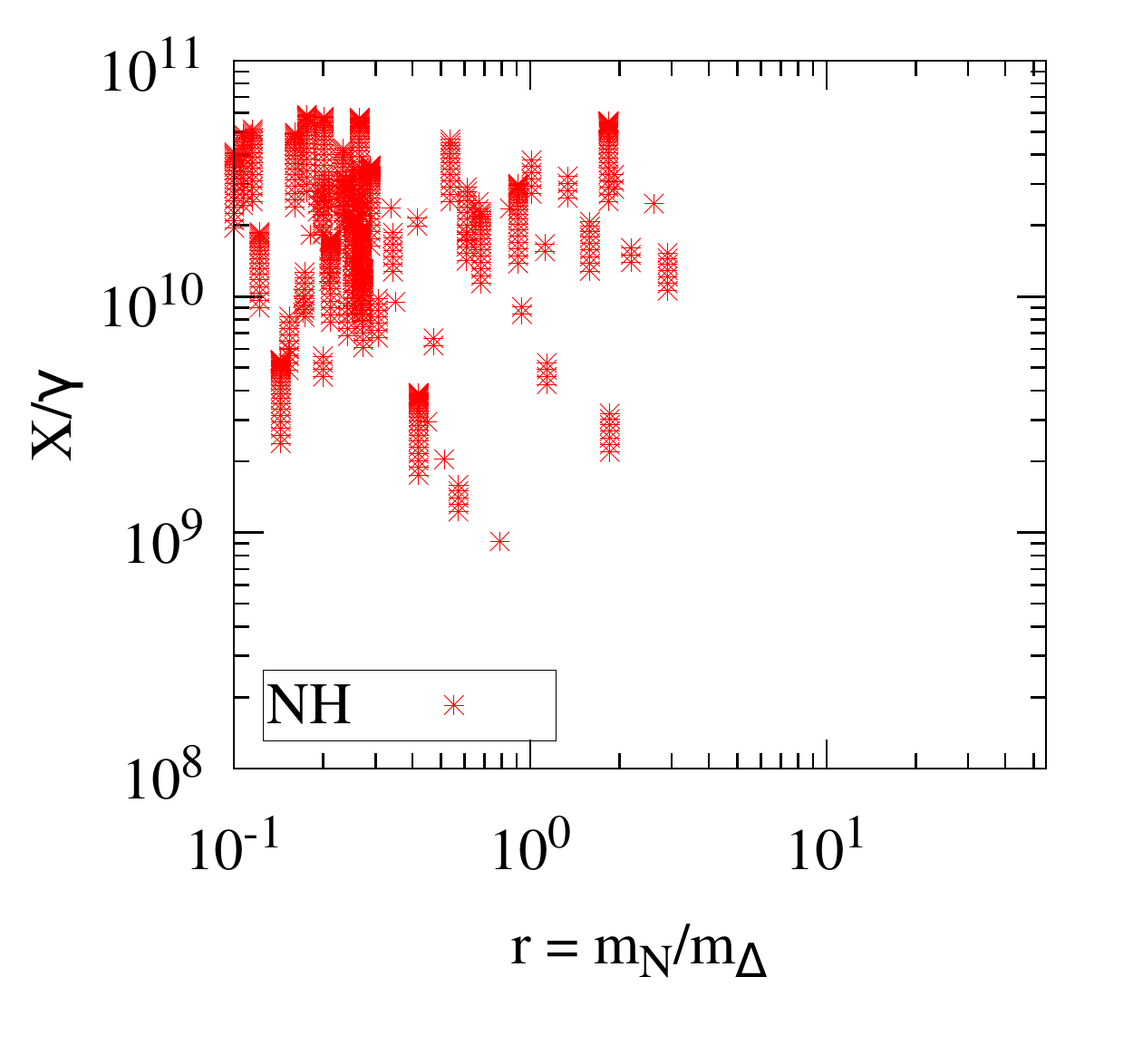,width=0.5\textwidth,clip=}&
\epsfig{file=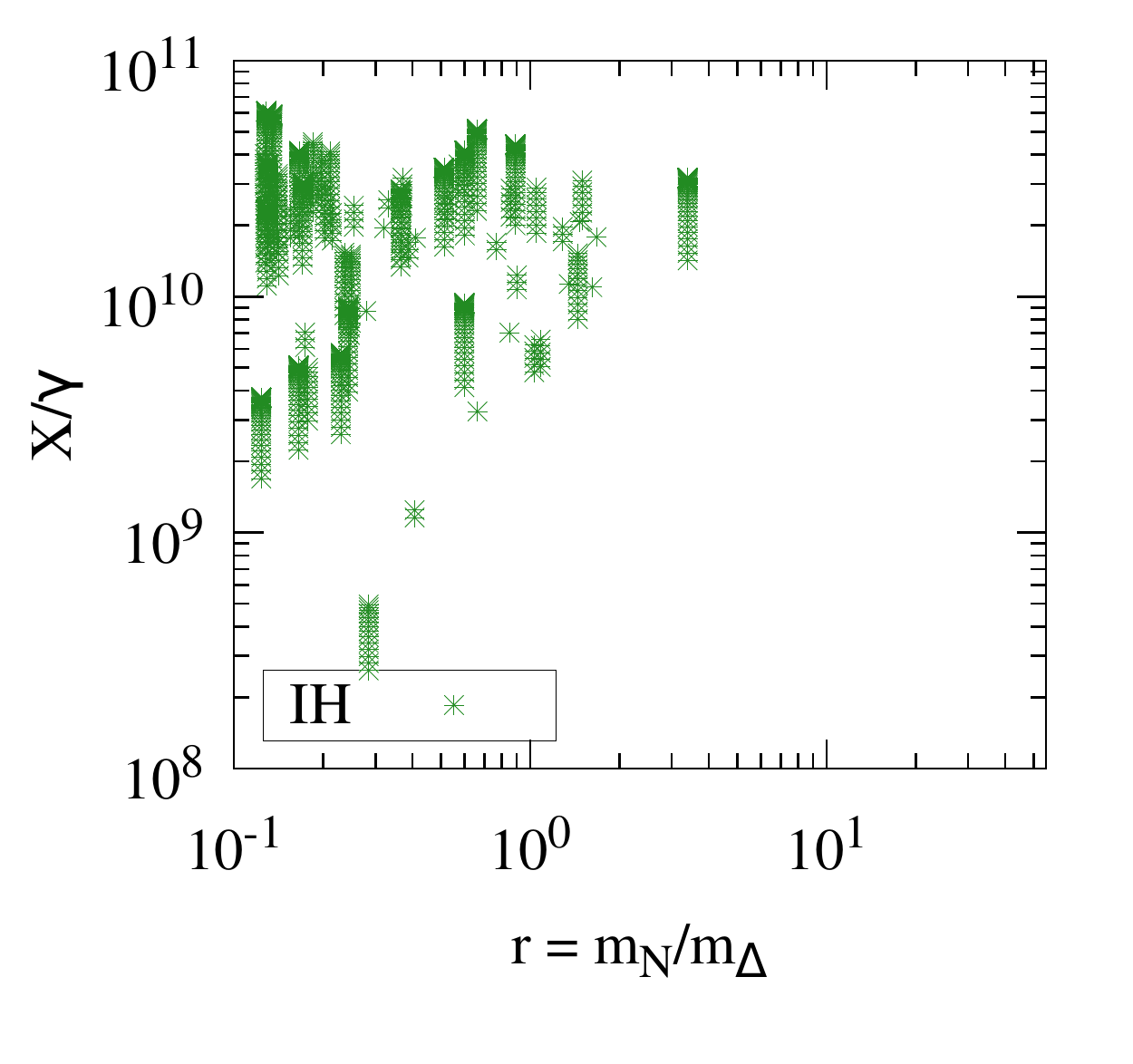,width=0.5\textwidth,clip=} \\
\end{tabular}
\caption{Allowed parameter space in $X/\gamma-r$ plane from constraints on neutrinoless double beta decay half-life and charged lepton flavour violation with type I+II seesaw.}
\label{fig7}
\end{figure}

\section{Results and Discussion}
\label{sec4}
We have studied the new physics contributions to neutrinoless double beta decay and charged lepton flavour violating processes $\mu \rightarrow e \gamma, \mu \rightarrow 3e$ within the framework of a TeV scale minimal left-right symmetric model. Keeping the right handed gauge boson masses within a few TeV such that they are accessible at particle colliders, we constrain the parameter space of the model by incorporating the latest experimental bounds on $0\nu \beta \beta$ and LFV amplitudes. Without adopting any specific structure of one of the seesaw mass matrices (considered in one of our earlier works), here we consider a general structure of type II seesaw mass matrix that can be diagonalised by a general orthogonal matrix. By varying the mixing angles of this orthogonal matrix and type II seesaw strength randomly, we calculate the right handed neutrino mass matrix as well as Dirac neutrino mass matrix for each of these choices. Choosing the best fit values of five light neutrino parameters, we randomly vary all other parameters affecting $0\nu \beta \beta$ and LFV and constrain them from experimental data. The other parameters which are being randomly varied are given in table \ref{param2}. The range of type II seesaw strength follows from the range for $X/\gamma$ given in equation \eqref{xgamma}. We also take into account the uncertainty in nuclear matrix elements involved in the calculation of $0\nu \beta \beta$ half-life. We show the total contribution to $0\nu \beta \beta$ half-life and LFV branching ratio as a function of lightest neutrino mass in figure \ref{fig3} and \ref{fig4} respectively. It can be seen from these plots that the existing experimental constraints on $0\nu \beta \beta$ half-life can not rule out any region of lightest neutrino mass $10^{-5}-10^{-1}$ eV, in such a general type I - type II seesaw scenario of MLRSM. However, as seen from figure \ref{fig4}, future observation of lepton flavour violating processes should be able to confirm some region of parameter space.

The interesting part of our results is the reopening of more regions of parameter space for $r = \frac{M_N}{M_{\Delta}}$ defined earlier. It can be seen from the plots shown in figure \ref{fig41}, \ref{fig5} and \ref{fig7} that this parameter can be larger than unity, implying that the doubly charged scalar masses can be as small as the heaviest right handed neutrino mass which can keep the scalar triplet masses well within the reach of LHC. This is in contrast to earlier results of \cite{ndbd00} showing the scalar triplet to be at least ten times heavier than the heaviest right handed neutrino and the more recent work \cite{ndbd102} where $r$ was shown to be close to unity for a very small range of lightest neutrino mass. As can be seen from the plot in figure \ref{fig5}, we can have $r \geq 1$ for almost all values of lightest neutrino mass in case of inverted hierarchy. For normal hierarchy, this gets restricted to a range $m_{\text{lightest}}/\text{eV} \in [3 \times 10^{-3}, 0.1]$. Although we have varied the masses of scalar triplets in the range 500 GeV to $\sqrt{4\pi}v_R$ shown in table \ref{param2} (where $v_R \approx 7.6$ TeV for $M_{W_R} = 3.5$ TeV), there is still room for lighter doubly charged scalar masses, if their branching ratio to leptons is not $100\%$, assumed by the LHC searches to put the exclusion limits \cite{hdlhc}. We also show the region of allowed parameter space in $X/\gamma-m_{\text{lightest}}$ and $X/\gamma-r$ planes in figure \ref{fig6} as well as \ref{fig7}. The range of $X/\gamma$ shown in these plots can be understood from the bound given in \eqref{xgamma} with our choices of parameters involved. 

With improving sensitivity at experiments like KamLAND-Zen and MEG resulting in their very recent updates on $0\nu\beta \beta$ half-life \cite{kamland2} and $\text{BR} (\mu \rightarrow e \gamma)$ \cite{MEG2}, the MLRSM particle spectrum has a high discovery potential at ongoing as well as future experiments looking for lepton flavour and lepton number violating decays. On the energy frontier, the ongoing LHC experiment may also come up with interesting signatures as it has the potential to scan $W_R$ masses upto around 6 TeV at 14 TeV centre of mass energy. This limit can go upto 35.5 TeV for future hadron colliders with 100 TeV centre of mass energy \cite{rizzo100}. Furthermore, linear lepton colliders like ILC and CLIC as well as electron-proton colliders like LHeC, FCC-eh have promising centre of mass energy reach to probe the TeV scale physics with high precision. All such planned future experimental setups should tremendously improve the discovery prospects of TeV scale MLRSM.

\begin{acknowledgments}
DB would like to express a special thanks to the Mainz Institute for Theoretical Physics (MITP) for its hospitality and support during the workshop \textit{Exploring the Energy Ladder of the Universe} where some part of this work was completed.
\end{acknowledgments}

\appendix
\label{appendix1}

\end{document}